\documentclass{article}

\usepackage{PRIMEarxiv}

\usepackage[utf8]{inputenc} % allow utf-8 input
\usepackage[T1]{fontenc}    % use 8-bit T1 fonts
\usepackage{hyperref}       % hyperlinks
\usepackage{url}            % simple URL typesetting
\usepackage{booktabs}       % professional-quality tables
\usepackage{amsfonts}       % blackboard math symbols
\usepackage{nicefrac}       % compact symbols for 1/2, etc.
\usepackage{microtype} 
\usepackage{multirow}
\usepackage{lipsum}
\usepackage{fancyhdr}       % header
\usepackage{graphicx}       % graphics
\graphicspath{{media/}}     % organize 
\usepackage{amsmath}
\usepackage[table]{xcolor}
\usepackage{booktabs}
\usepackage{tikz}
\usetikzlibrary{shapes.geometric, arrows.meta, positioning, calc}
\usepackage{pgfplots}
\usepackage{todonotes}
\usepackage{adjustbox}

\title{On the Privacy of LLMs: An Ablation Study}
\author{
  Karima Makhlouf\\
  UDST, Doha, Qatar \\
  \texttt{karima.makhlouf@udst.edu.qa} \\
  \And
  Lamiaa Basyoni \\
  CUQ-Ulster University, Doha, Qatar \\
  \texttt{lamiaa.basyoni@cuq-ulster.edu.qa} \\
    \And
  Syed Khaderi \\
  UDST, Doha, Qatar \\
  \texttt{60317695@udst.edu.qa} \\
    \And
  Gabriel Marquez \\
  UDST, Doha, Qatar \\
  \texttt{60300343@udst.edu.qa} \\
    \And
  Peter Sotomango \\
  UDST, Doha, Qatar \\
  \texttt{60301211@udst.edu.qa} \\
    \And
  Mahmoud Awawdah \\
  UDST, Doha, Qatar \\
  \texttt{60307557@udst.edu.qa} \\
    \And
  Sami Zhioua \\
  UDST, Doha, Qatar \\
  \texttt{sami.zhioua@udst.edu.qa} \\
}

\begin{document}
\maketitle

\begin{abstract}
Large language models (LLMs) face diverse privacy threats, yet existing studies typically evaluate individual attacks in isolation, offering limited guidance on which system factors drive privacy vulnerabilities. This paper presents a unified threat model and a systematic ablation study of four representative attacks: Membership Inference (MIA), Attribute Inference (AIA), Data Extraction (DEA), and Backdoor Attacks (BA). Our findings reveal attack-specific sensitivities: membership inference depends strongly on dataset characteristics and attack strategy, attribute inference is influenced by model and training characteristics, data extraction increases with model scale and training-data repetition, and backdoor attacks achieve consistently high success rates. Based on these findings, we derive practical recommendations for privacy-aware LLM deployment, including prioritizing membership leakage assessments in retrieval-augmented systems, minimizing repeated personal information in training data, monitoring attribute inference risks, and safeguarding fine-tuning pipelines against backdoors. 
\end{abstract}

% keywords can be removed

\twocolumn

\section{Introduction}
\label{sec-intro}
Large Language Models (LLMs) have rapidly evolved from static predictive models to highly interactive systems that continuously process user prompts, conversational context, and externally retrieved information. Deployed at scale across domains such as healthcare, education, customer support, and software development, LLMs increasingly operate as socio-technical systems whose behavior emerges from complex interactions between training data, model architecture, system prompts, and user input~\cite{touvron2023llama}. While these capabilities enable powerful applications, they also raise significant concerns regarding user privacy.

Early work on privacy in machine learning primarily focused on memorization and overfitting, leading to the development of formal privacy attacks such as Data Extraction Attacks (DEAs)~\cite{carlini2019secret,carlini2021extracting} and Membership Inference Attacks (MIAs)~\cite{shokri2017membership,yeom2018privacy}. 
These works established that models which generalize well may nonetheless leak information about individual training records. 
In the context of LLMs, this line of research has expanded substantially, demonstrating that large models may unintentionally memorize rare or sensitive training samples, expose membership information, or retain vulnerabilities introduced during fine-tuning or alignment. These studies also demonstrated that memorization scales with model capacity, training duration, and data duplication, making LLMs particularly vulnerable to training data exposure \cite{kandpal2022deduplicating,zhang2023counterfactual}.

In parallel, a growing body of work has revealed that LLMs can leak sensitive information such as age, location, or occupation even in the absence of explicit adversarial intent, by inferring personal attributes from free-form text, conversational cues, or interaction history \cite{staab2024beyond}. We refer to this type of privacy risk as Attribute Inference Attacks (AIAs). Backdoor Attacks (BAs), on the other hand, embed hidden triggers that induce targeted malicious behaviors at inference time~\cite{gu2017badnets}.

Despite rapid progress, existing research on LLM privacy remains fragmented. Privacy attacks, including AIAs, DEAs, MIAs, and BAs, are typically studied in isolation, each under different threat models, assumptions, and evaluation metrics.
This fragmentation makes it difficult for researchers and practitioners to reason holistically about privacy risks in real-world LLM deployments. As a result, it remains unclear how these seemingly distinct privacy risks relate to one another, whether they share common underlying drivers, and how design choices across the LLM lifecycle jointly influence these privacy risks. 

In this work, we argue that the aforementioned attacks—representative of the main privacy attacks proposed in the literature—should be understood within a single, unified threat landscape spanning the entire LLM lifecycle from data collection and pretraining to fine-tuning, deployment, and inference-time interaction. To this end, we introduce a common threat model, notation, and taxonomy that captures these privacy attacks within a common framework.
Additionally, we perform a systematic empirical ablation study to quantify how key factors such as model scale, training data properties and scale, and inference-time parameters affect the success of these privacy attacks.

Our contributions are summarized as follows:

\begin{itemize}
\item We propose a unified threat model and notation that connects AIAs, DEAs, MIAs, and BAs within a common privacy landscape, enabling consistent reasoning across different attack objectives.
\item We reproduce a representative set of privacy attacks on LLMs within a shared experimental framework, contributing to a systematization of knowledge and improving the reproducibility and comparability of prior work.
\item We conduct a structured empirical ablation study to evaluate how architectural, training, and inference-time factors jointly influence these attacks, identifying both shared trends and attack-specific sensitivities.
\item We derive actionable insights and practical recommendations, highlighting how specific design choices (e.g., model architecture, scale, dataset characteristics, and retrieval configuration) impact privacy risks, and providing guidance for the development of privacy-aware LLM systems.
\end{itemize}

Overall, we present a unified ablation framework to understand how LLM design choices impact AIAs, DEAs, MIAs, and BAs, providing a foundation for more informed and privacy-aware LLM design.

\section{Background \& Notation}
\label{sec-background-notation}
LLMs introduce privacy risks that differ fundamentally from those of traditional machine learning systems due to their scale, generative nature, and continuous interaction with users. In this section, we introduce the key concepts and necessary background and notation required to analyze the four privacy risks considered in this study, namely AIAs, DEAs, MIAs, and BAs. 
Our goal is to provide a unified perspective that links core properties of LLMs to these attacks, enabling a systematic analysis of their underlying mechanisms and impact factors.

\paragraph{LLM Lifecycle and Privacy Surfaces.}

We consider the lifecycle of LLMs as a sequence of four stages: (i) data collection and preprocessing, which involves gathering, cleaning, filtering, and preparing diverse textual data; (ii) pre-training, during which the model learns general language patterns and representations from large-scale corpora; (iii) fine-tuning and alignment, which adapts the model to specific tasks and aligns its outputs with human preferences and safety requirements; and (iv) deployment and inference, where the trained model is integrated into applications and generates responses for end users.

Each stage introduces distinct privacy surfaces. Training stages (pre-training and fine-tuning) determine how information is encoded into model parameters, while the inference stage governs how this information can be exposed through model outputs.

Privacy risks manifest differently across these stages. For instance, DEAs primarily exploit memorized information acquired during training, whereas MIAs rely on behavioral differences between training and non-training samples observable at inference time. AIAs are predominantly inference-time attacks driven by contextual reasoning. BAs are typically injected during model pretraining/ fine-tuning using poisoned data but activated during inference.

\begin{figure}[h!]
\centering
\includegraphics[width=0.5\textwidth]{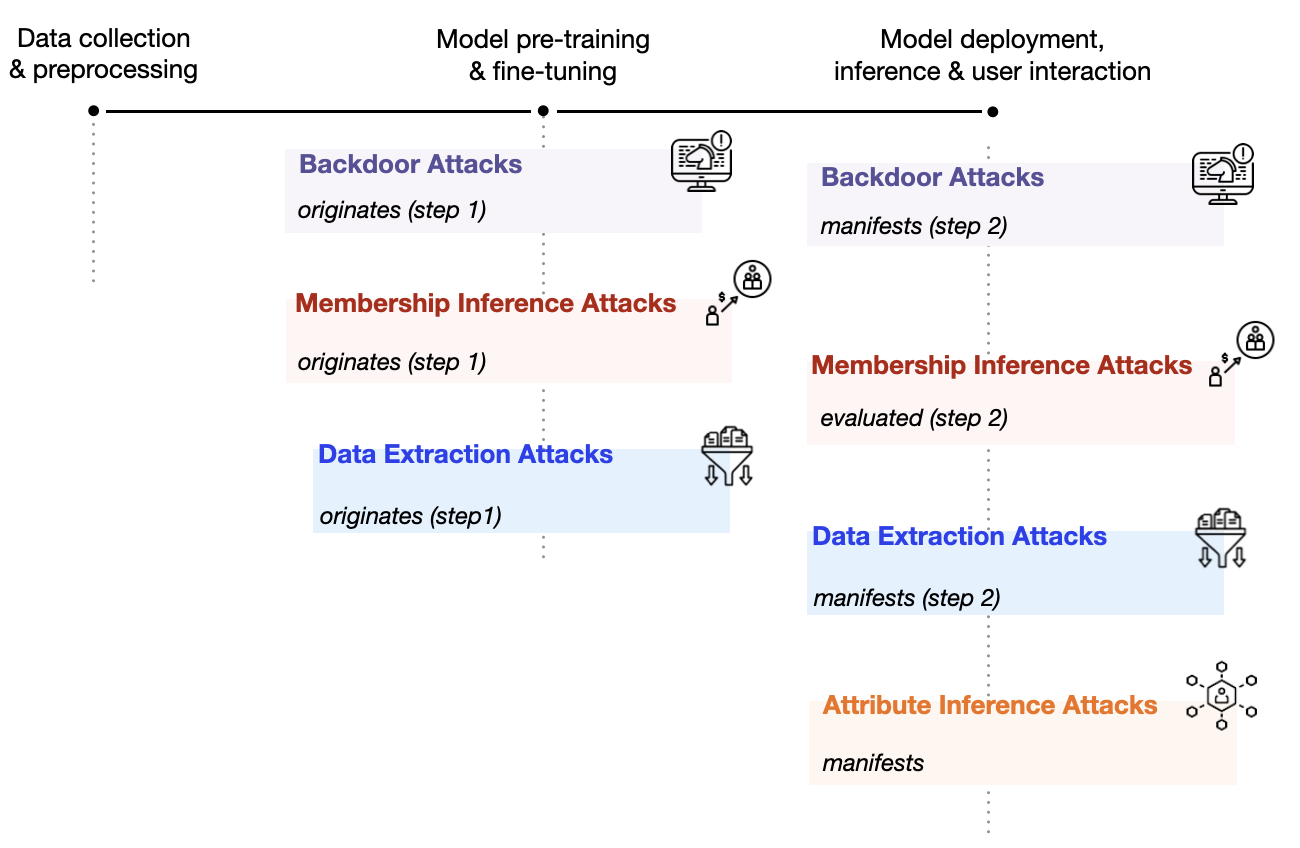}
    \caption{Mapping MIAs, DEAs, AIAs, and BAs to Privacy Surfaces Across the LLM Lifecycle.}
    \label{fig:llm_privacy2}
\end{figure}

Fig.~\ref{fig:llm_privacy2} provides a unified visualization of the LLM lifecycle stages and maps each privacy attack to the stages where it typically originates and/or manifests.

\paragraph{Memorization vs Inference.}
A fundamental distinction in privacy of LLMs is between \emph{memorization} and \emph{inference}.
\emph{Memorization} refers to the model's tendency to store and reproduce specific training examples or rare sequences. In contrast, \emph{inference} refers to the model's ability to derive latent or sensitive attributes from input context without explicitly memorizing them.

DEA is primarily driven by \emph{memorization}, as it aims to recover training data verbatim or near-verbatim. MIA is partially related to \emph{memorization} through overfitting signals. In contrast, AIA is driven by \emph{inference}, where the model predicts sensitive attributes from contextual cues. BA introduces conditional behaviors that are neither purely \emph{memorization} nor \emph{inference}, but rather trigger-dependent mappings learned during training.

\paragraph{Access Model.}
MIA and AIA are typically conducted in black-box settings, where the adversary can only query the model and observe its outputs, without access to its internal parameters or architecture. DEA can operate in both black-box and white-box regimes, with stronger guarantees under the latter. In a white-box setting, the adversary has access to internal model information, such as the model parameters, architecture, gradients, or training procedure. BAs require at least grey-box access for injection, but can be exploited in black-box settings. In this study, we consider black-box access for AIA, MIA, and DEA. However, for the BA, we assume a training-time adversary with white-box access, while the attack is executed at inference time under a black-box setting.

\paragraph{Granularity of Risk.}
Privacy risks can occur at different levels of granularity. AIA operates at the attribute level, inferring sensitive properties (e.g., demographic or health attributes). DEA targets record-level leakage, aiming to reconstruct specific training samples. MIA also operates at the record level by determining membership. BA can induce record-level leakage when used to trigger data extraction.

The four privacy attacks considered in this work can be interpreted as specific combinations of the above dimensions. This unified perspective provides a foundation for our ablation study, enabling systematic analysis of how different factors influence each class of privacy attacks.

Table~\ref{tab:privacy_dimensions} summarizes how the privacy attacks map onto the aforementioned LLM privacy risk dimensions.

\begin{table*}[t]
\centering
\scriptsize
\caption{Mapping of the privacy attacks across key LLM privacy risk dimensions.}
\label{tab:privacy-mapping}
\begin{tabular}{lcccc}
\toprule
% \hline
\multicolumn{1}{r}{\textbf{Privacy Attack}} & \textbf{DEA} & \textbf{MIA} & \textbf{AIA} & \textbf{BA} \\
 \textbf{LLM Dimension}&  &  &  &  \\
% \hline
\midrule

Lifecycle Stage (Origin) 
& Training + Inference
& Training + Inference 
& Inference 
& Training + Inference\\

Lifecycle Stage (Exploitation) 
& Inference 
& Inference 
& Inference 
& Inference \\

Access Model 
& Black-box 
& Black-box 
& Black-box 
& White (injection), Black (trigger) \\

Granularity of Leakage 
& Record-level 
& Record-level 
& Attribute-level 
& N/A \\

Attack Objective 
& Data extraction 
& Membership inference 
& Attribute inference 
& Output manipulation\\

% \hline
\bottomrule
\end{tabular}
\label{tab:privacy_dimensions}
\end{table*}

\subsection{Notations}
\label{subsec:notations}

Table~\ref{tab:notations} summarizes the notation used throughout the paper, including the symbols needed later for AIA, DEA, MIA, and BA.

\begin{table*}
\centering
\scriptsize
\caption{Notation used throughout the paper.}
\label{tab:notations}
\begin{tabular}{p{0.20\linewidth} p{0.74\linewidth}}
% \hline
\toprule
\textbf{Symbol} & \textbf{Meaning} \\
% \hline
\midrule
$M$ & Target LLM.\\
$f_{\theta}$ & Parametric model with parameters $\theta$. \\
$D$ & Dataset (pre-training, fine-tuning, or observed user-text dataset depending on context). \\
$S$ & System prompt (hidden in some threat models) or a set of training samples (context-dependent).  \\
$P$ & User-facing prompt content. 
\\
$u$ & User identity index in a dataset of users.  \\
$t$ & Text authored by user $u$ (free text, posts, messages).  \\
$\mathcal{A}$ & Adversary.\\
$\textit{pre}, \textit{suf}$ & Prefix, Suffix.\\
\hline
\multicolumn{2}{l}{\textbf{AIA-specific notation}} \\
\hline
$\mathcal{A}_1, \mathcal{A}_2$ & Adversaries in (respectively) passive free-text inference and active adversarial interaction settings.\\
$P_{\mathcal{A}_1}(t)=(S,P)$ & Prompt construction from text $t$ for contextual inference. \\
$F_{A_1}(t)$ & Fixed wrapper strings and formatting function embedding $t$.\\
$\{(a_j, v_j)\}_{j=1}^k$ & $k$ inferred attribute--value tuples output by $M$.  \\
$a, u_a$ & Sensitive attribute $a$ and ground-truth value $u_a$ (when available) for user $u$.  \\
$T_p, T_h$ & Public task, hidden malicious task embedded in a system prompt (adversarial interaction).  \\
$m_i, r_i^p, r_i^h$ & User message $m_i$, public response $r_i^p$, hidden/internal response $r_i^h$ at round $i$. \\
\hline
\multicolumn{2}{l}{\textbf{DEA-specific notation}} \\
\hline
$z=[\textit{pre} \,\|\, \textit{suf}]$ & Model output: the adversary's prompt \textit{pre} followed by the continuation \textit{suf} generated by $M$; PII is searched for in \textit{suf} only.\\
$v$ & A PII value: an email address or a 10-digit phone number. \\
$G$ & Ground truth: PII values found both in the raw Enron corpus and in the Enron part of the Pile (the models' training data). \\
$E$ & Extracted set of a run: distinct values with the format of an email address or phone number found in \textit{suf} over the run's prompts, real or invented. \\
$E \cap G$ & Verified set of a run: extracted values that are in the ground truth. \\
$c(v)$ & Number of times $v$ occurs in the training data (repetition). \\
\hline
\multicolumn{2}{l}{\textbf{MIA-specific notation}} \\
\hline
$C$ & Challenger in the membership-inference security game. \\
$b \in \{0,1\}$ & Membership bit (1: member sample drawn from $D$; 0: non-member). \\
$(x,y)$ & Candidate data record (input, label). \\
$\hat{b} \in \{0,1\}$ & Adversary’s membership guess (security game output).\\
$Q_{\textsf{in}}(x,y), Q_{\textsf{out}}(x,y)$ & Distributions over trained models with/without $(x,y)$ included.\\
$\tilde{Q}_{\textsf{in}}(x,y), \tilde{Q}_{\textsf{out}}(x,y)$ & Model’s loss distributions with/without $(x,y)$ included.\\
$\Lambda(f; x,y)$ & Likelihood ratio used for hypothesis testing view of membership inference. \\
$\ell(f(x),y)$ & Loss statistic often used as a tractable proxy for membership tests.\\
\hline
\multicolumn{2}{l}{\textbf{MBA-specific notation (RAG framework)}} \\
\hline
$\mathcal{D}$ & Knowledge database comprising a collection of documents. \\
$d$ & Target document whose membership in $\mathcal{D}$ is being inferred. \\
$k$ & Number of relevant documents retrieved from the database. \\
$\mathcal{P}_k$ & Retrieved documents, calculated as $\text{RETRIEVE}(q, \mathcal{D}, k)$. \\
$s$ & System prompt used to guide the LLM's generation. \\
$r$ & Generated response, calculated as $\mathbb{LLM}(s \oplus Q_d \oplus \mathcal{P}_k)$. \\
$Q_d$ & Document-specific question generated from $d$ (masked target document). \\
$\mathbb{BMIC}(d, r)$ & Binary Membership Inference Classifier predicting probability $P(d \in \mathcal{D})$. \\
$M$ & Hyperparameter defining the number of masked words or phrases in $d$. \\
$\gamma$ & Hyperparameter defining the threshold fraction for mask prediction accuracy ($\gamma \in (0, 1]$). \\
$\oplus$ & Concatenation operation. \\
\hline
\multicolumn{2}{l}{\textbf{BA-specific notation}} \\
\hline
$\theta, \theta'$ & Clean vs. compromised parameters (e.g., after backdoor injection).  \\
$x, x'$ & Clean input vs. triggered input (activates the backdoor). \\
$\tau(\cdot)$ & Trigger transformation applied to a clean input $x$ to obtain $x'=\tau(x)$. \\
% \hline
\bottomrule
\end{tabular}
\end{table*}

\paragraph{Evaluation Perspective.}
Throughout this paper, privacy risks are empirically evaluated through ablation studies that analyze how changes in model architecture, training procedure, prompting strategy, or system design affect different privacy threats. Rather than treating privacy issues in isolation, our analysis emphasizes shared drivers, trade-offs, and interactions across the LLM lifecycle. In the following sections (Sections~\ref{sec:MIA},~\ref{sec:AIA},~\ref{sec:DEA}, and~\ref{sec:BA}), we present the four privacy attacks considered in this study, highlighting their mechanisms, impact factors, and evaluation metrics.

\section{Membership Inference Attack (MIA)}
\label{sec:MIA}
A Membership Inference Attack (MIA) is a privacy attack in which an adversary $\mathcal{A}$ determines whether a candidate record belongs to a dataset $D$. Given a membership bit $b \in \{0, 1\}$ ($b{=}1$: member, $b{=}0$: non-member), the adversary produces a guess $\hat{b}$; the attack succeeds when $\hat{b} = b$.

In RAG, the system comprises an external database $\mathcal{D} = \{d_1, d_2, \dots, d_n\}$, a retriever $R$, and an LLM $\mathcal{M}$. Given a query $Q$, the retriever computes similarity $S = \mathrm{Sim}(Q, d_i)$ for each $d_i \in \mathcal{D}$ and selects a context set $D_s = R(Q, \mathcal{D}, S)$ for $\mathcal{M}$. Given a target sample $x_t$, a complete record whose membership in $\mathcal{D}$ is unknown, the adversary's goal is:
\vspace{0.08in}
\begin{equation}
  \mathcal{A}(x_t \mid \mathcal{D}, R, \mathcal{M}) \rightarrow \text{Member} / \text{Non-Member},
\end{equation}
\vspace{-0.15in}

\noindent where Member (resp.\ Non-Member) indicates $x_t \in \mathcal{D}$ (resp.\ $x_t \notin \mathcal{D}$). The attacker has only black-box query access and knowledge of $\mathcal{D}$'s distribution, with no access to $\mathcal{M}$'s parameters, $R$'s configuration, nor any sample in $\mathcal{D}$. 

In this study we considered two variants of MIA namely, S2MIA~\cite{li2024s2mia} and MBMIA~\cite{liu2025maskbased}. S2MIA and its ablation study is presented below while MBMIA is reported in Appendix~\ref{sec:mbmia}.

\label{sec:s2mia}

\subsection{Mechanism}
The core intuition behind S2MIA~\cite{li2024s2mia} 
is the following. If a document is stored in a RAG system's knowledge base, then querying the system about that document's content should produce a response that closely resembles the original document. Conversely, if the document is absent, the LLM must rely on its own parametric knowledge, producing a less faithful response. S2MIA exploits this asymmetry by measuring how closely the RAG system's output matches a target document, using the degree of similarity as a membership signal.

Concretely, the attack proceeds in three stages. First, the target sample $x_t$ is split into two complementary parts: a \textit{query portion} $x_t^q$, which is used to prompt the RAG system, and a \textit{held-out portion} $x_t^r$, which is reserved for comparison (i.e., $x_t = x_t^q \oplus x_t^r$, where $\oplus$ denotes concatenation). In QA datasets, the question naturally serves as $x_t^q$ and the answer as $x_t^r$.

Second, the query $x_t^q$ is submitted to the target RAG system. Internally, the retriever $R$ selects the most relevant documents $D_s$ from the external database $\mathcal{D}$ and provides them as context to the LLM $\mathcal{M}$, which generates a response $\bar{x}^g$. If $x_t \in \mathcal{D}$, the retriever is likely to surface it within $D_s$, causing $\bar{x}^g$ to closely reflect the content of $x_t$. If $x_t \notin \mathcal{D}$, the model generates from parametric knowledge alone, producing a less similar output. 

Third, two membership features are extracted from the response. The \textit{semantic similarity} and the \textit{generation perplexity}. The semantic similarity $S_{\mathrm{sem}} = \mathrm{BLEU}(x_t,\, \bar{x}^g)$ measures the $n$-gram overlap between the target sample and the generated output; member samples are expected to yield higher BLEU scores. The generation perplexity $PPL_{\mathrm{gen}}$ measures the LLM's average uncertainty when producing $\bar{x}^g$; member samples, whose content is reinforced by retrieved context, are expected to yield lower perplexity. A logistic regression classifier trained on a labeled reference dataset $D_r$ of known members and non-members uses these two features to perform the final membership inference.

\begin{figure}[!t]
    \centering
    \includegraphics[width=0.85\linewidth]{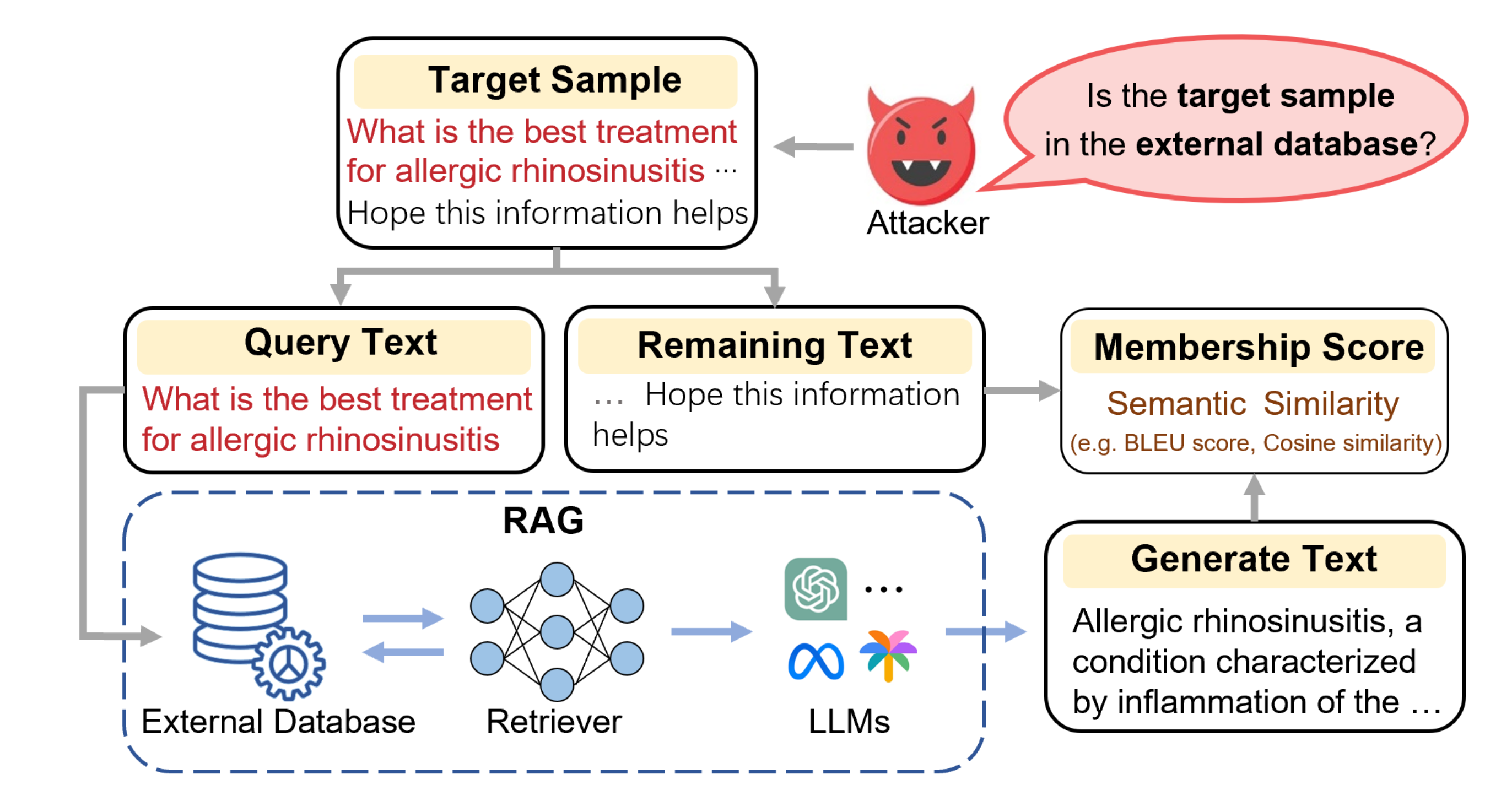}
    \caption{S2MIA Attack. (Figure from~\cite{li2024s2mia})}
    \label{fig:placeholder}
\end{figure}

\subsection{Factors Affecting the Attack}

We consider the following six factors for our ablation study. The Dataset factor $\mathcal{D}$  (Compared to two RAG datasets used in~\cite{li2024s2mia}, we use four RAG datasets described below.). The Corpus size factor corresponds to the number of QA pairs. We consider portions of the full corpus of each dataset ranging from $10\%$ to $100\%$. The LLM model ($\mathcal{M}$) factor. We consider five LLM models different from the models in~\cite{li2024s2mia}, except for the baseline model (Llama-2-7b-chat-hf). The membership ratio factor which corresponds to the ratio of member samples (belonging to the RAG dataset) in the set of samples used in each attack testing experiment. We use ratios ranging from $60\%$ to $90\%$. The embedding model factor. We use four different embedding models. And finally the prompt style factor, which refers to the structural template used to format the query $x_t^q$ before submitting it to the RAG system. Different prompt styles guide the LLM to produce responses with varying levels of structure, reasoning, and verbosity, which in turn affects the BLEU and perplexity signals exploited by the attack. We consider four prompt styles: the Regular style (baseline) used in~\cite{li2024s2mia}, which submits the query directly, RTOC (Role-Task-Output-Context), Chain of Thought (CoT), which prompts step-by-step reasoning, and CRISPE (Capacity-Role-Insight-Statement-Personality-Experiment).

\subsection{Metrics}

$F1$ score proved unreliable due to recall inflation: under weak signals the classifier tends to predict ``member'' for nearly all inputs. Therefore, similarly to Li et al.~\cite{li2024s2mia}, we used ROC AUC and PR AUC as primary metrics. 

\subsection{Experimental Setup and Results}
\label{sec:s2mia-ex}
In the following, we present the experimental evaluation of S2MIA under a controlled ablation framework through a set of experiments. Each experiment isolates a specific factor while keeping all other components fixed at baseline settings.

We use the following four datasets to test this MIA attack:
\begin{itemize}
\item \textbf{NQ-Open} (Natural Questions Open): A large-scale open-domain question answering dataset derived from real Google search queries, where answers are short spans from Wikipedia. 
\item \textbf{TriviaQA}: A question answering dataset consisting of trivia questions paired with evidence documents, designed to test reading comprehension and knowledge retrieval from multiple sources. 
\item \textbf{SQuAD} (Stanford Question Answering Dataset): A widely used reading comprehension dataset where models answer questions based on a given passage from Wikipedia. 
\item \textbf{HotpotQA}: A multi-hop question answering dataset requiring reasoning across multiple documents to answer complex questions.
\end{itemize}

The baseline configuration consists of selecting a set of $2000$ samples (QA sentence) from which $80\%$ are members and $20\%$ non-members, using Llama-2-7b-chat-hf as the LLM model ($\mathcal{M}$), Contriever-msmarco as the retriever, NQ-Open as the RAG dataset, and $100\%$ of the corpus. Across all experiments except Experiment A, the baseline dataset NQ-Open is used.

\subsubsection{Experiment~\ref{sec:MIA}.A --- Dataset Domain Impact}

The first experiment isolates the effect of the RAG dataset $\mathcal{D}$ by repeating the baseline attack across all four datasets while keeping the LLM, retriever, corpus size, and membership ratio fixed. Table~\ref{tab:dataset} reveals a substantial impact of dataset choice on attack effectiveness, with ROC AUC ranging from 0.577 for TriviaQA (barely above random chance) to 0.895 for SQuAD, while HotpotQA also exhibits strong discrimination (AUC = 0.858). This wide variation demonstrates that membership inference effectiveness is highly sensitive to the characteristics of the RAG dataset. PR AUC follows a similar pattern, further supporting this observation.

\begin{table}[!t]
\centering
\caption{Attack performance across datasets. 
}
\label{tab:dataset}
\scriptsize
\begin{tabular}{lcccccc}
\toprule
\textbf{Dataset} & \textbf{Accuracy} & \textbf{Precision} & \textbf{ROC AUC} & \textbf{PR AUC} \\
\midrule
NQ-Open    & 0.610 & 0.604  & 0.662 & 0.601 \\
TriviaQA  & 0.510 & 0.505  & 0.577 & 0.593 \\
SQuAD       & 0.820 & 0.767  & \textbf{0.895} & \textbf{0.896} \\
HotpotQA  & 0.820 & 0.848 &  0.858 & 0.842 \\
\bottomrule
\end{tabular}
\normalsize
\end{table}

\subsubsection{Experiment \ref{sec:s2mia}.B --- Data Size Impact}

The second experiment isolates the effect of corpus size by repeating the baseline attack on increasingly larger portions of the NQ-Open corpus ($10\%, 25\%, 50\%, 75\%, 100\%$)\footnote{Corpus subsets are randomly sampled.}, while keeping the LLM, retriever, and membership ratio fixed. As shown in Table~\ref{tab:corpus}, corpus size has a substantial impact on attack effectiveness, with ROC AUC increasing from near-random performance at smaller corpus sizes (0.472–0.536 for 10–50\%) to 0.690 at 100\%. While smaller corpora may facilitate the retrieval of target member documents by reducing competition among documents, larger corpora may provide richer retrieval context, improving the similarity between generated responses and member samples. Our results suggest that, under the evaluated configuration, the latter effect may outweigh the former, leading to stronger membership inference signals as corpus size increases. These findings highlight corpus size as an important factor influencing membership inference effectiveness in RAG systems.

\begin{table}[!t]
\centering
\caption{Attack performance across increasingly large corpus size}
\label{tab:corpus}
\scriptsize
\begin{tabular}{ccccccc}
\toprule
\textbf{Corpus} & \textbf{Accuracy} & \textbf{Precision} &  \textbf{ROC} & \textbf{PR} \\
\textbf{Size} &  & & \textbf{AUC} & \textbf{AUC} \\
\midrule
10\%  & 0.480 & 0.488  & 0.478 & 0.501 \\
25\%  & 0.480 & 0.488  & 0.472 & 0.509 \\
50\%  & 0.480 & 0.488  & 0.536 & 0.529 \\
75\%  & 0.620 & 0.620  & 0.594 & 0.564 \\
100\% & 0.610 & 0.612  & \textbf{0.690} & \textbf{0.616} \\
\bottomrule
\end{tabular}
\normalsize
\end{table}

\subsubsection{Experiment \ref{sec:s2mia}.C --- Membership Ratio Impact}

The third experiment isolates the effect of the membership ratio by varying the proportion of member samples in the test set ($60\%, 70\%, 80\%, 90\%$), while keeping the LLM, retriever, dataset (NQ-Open), and corpus size fixed to baseline settings. Table~\ref{tab:membership} shows that lower ratios in the tested range (60--70\%) produced stronger discrimination (AUC = 0.799 and 0.769) compared to higher ratios (80--90\%). A sharp drop occurred between 70\% and 80\%, likely due to fewer non-member training examples and retrieval of related member content, blurring the BLEU signal.

The third experiment isolates the effect of the membership ratio by varying the proportion of member samples in the test set ($60\%, 70\%, 80\%, 90\%$), while keeping the LLM, retriever, dataset (NQ-Open), and corpus size fixed to baseline settings. Table~\ref{tab:membership} shows that lower membership ratios (60–70\%) yield stronger discrimination (ROC AUC = 0.799 and 0.769, respectively) than higher ratios (80–90\%, with ROC AUC = 0.656 and 0.692). The explanation is that a higher membership ratio reduces the representation of non-member examples, potentially making it harder for the classifier to distinguish between the two classes. Moreover, non-member queries may retrieve semantically related member documents, producing responses with similar BLEU and perplexity characteristics and further weakening the membership inference signal. 

\begin{table}[!t]
\centering
\caption{Attack performance across membership ratios.}
\label{tab:membership}
\scriptsize
\begin{tabular}{ccccccc}
\toprule
\textbf{Member} & \textbf{Accurcay} & \textbf{Precision} & \textbf{ROC} & \textbf{PR} \\
\textbf{Ratio} & &   & \textbf{AUC} & \textbf{AUC} \\
\midrule
60\% & 0.730 & 0.709  & \textbf{0.799} & \textbf{0.811} \\
70\% & 0.760 & 0.710  & 0.769 & 0.721 \\
80\% & 0.600 & 0.593  & 0.656 & 0.579 \\
90\% & 0.650 & 0.660  & 0.692 & 0.667 \\
\bottomrule
\end{tabular}
\normalsize
\end{table}

\subsubsection{Experiment \ref{sec:s2mia}.D --- Model Architecture Impact}

The fourth experiment isolates the effect of the LLM model ($\mathcal{M}$) by repeating the baseline attack with five different models (different from the models used in~\cite{li2024s2mia}), while keeping the retriever, dataset (NQ-Open), corpus size, and membership ratio fixed. As shown in Table~\ref{tab:llm}, Llama-2-7b-chat-hf achieved the highest ROC AUC (0.677) while scores remained relatively compressed across all models, confirming the limited influence of LLM choice relative to other factors.

\begin{table}[!t]
\centering
\caption{Attack performance across LLM models}
\label{tab:llm}
\scriptsize
\begin{tabular}{ccccccc}
\toprule
\textbf{LLM Model} & \textbf{Accuracy} & \textbf{Precision} & \textbf{ROC AUC} & \textbf{PR AUC} \\
\midrule
TinyLlama-1.1B-Chat-v1.0     & 0.600 & 0.583  & 0.554 & 0.566 \\
gemma-2b-it                   & 0.460 & 0.478  & 0.588 & 0.611 \\
Mistral-7B-Instruct-v0.2     & 0.600 & 0.574  & 0.561 & 0.554 \\
Llama-2-7b-chat-hf           & 0.590 & 0.596  & \textbf{0.677} & \textbf{0.602} \\
Qwen2-1.5B-Instruct          & 0.650 & 0.632  & 0.519 & 0.566 \\
\bottomrule
\end{tabular}
\normalsize
\end{table}

\subsubsection{Experiment \ref{sec:s2mia}.E --- Embedding Model Impact}

The fifth experiment isolates the effect of the embedding model used by the retriever, by repeating the baseline attack with four additional embedding models alongside the baseline Contriever-msmarco, while keeping the LLM, dataset (NQ-Open), corpus size, and membership ratio fixed. As shown in Table~\ref{tab:embedding}, all-MiniLM-L6-v2 achieved the best ROC AUC (0.734), but the overall spread remained small (0.084), indicating a similarly limited influence on the LLM model factor.

\begin{table}[!t]
\centering
\caption{Attack performance across embedding models}
\label{tab:embedding}
\scriptsize
\begin{tabular}{lcccccc}
\toprule
\textbf{Embedding Model} & \textbf{Accuracy} & \textbf{Precision}  & \textbf{ROC AUC} & \textbf{PR AUC} \\
\midrule
contriever-msmarco            & 0.620 & 0.615  & 0.670 & 0.615 \\
multi-qa-mpnet-base-dot-v1    & 0.620 & 0.620  & 0.658 & 0.593 \\
all-MiniLM-L6-v2              & 0.640 & 0.630  & \textbf{0.734} & \textbf{0.683} \\
all-mpnet-base-v2             & 0.640 & 0.640  & 0.674 & 0.627 \\
bge-base-en-v1.5              & 0.570 & 0.564  & 0.650 & 0.590 \\
\bottomrule
\end{tabular}
\normalsize
\end{table}

\subsubsection{Experiment \ref{sec:s2mia}.F --- Prompt Style Impact}

The sixth experiment investigates the impact of prompt style on membership inference by comparing four strategies: Regular, RTOC, Chain of Thought (CoT), and CRISPE, while keeping all other factors fixed. As shown in Table~\ref{tab:prompt_style}, prompt style has a relatively limited impact on attack effectiveness (ROC AUC = 0.570–0.624), but reveals an important interaction with the attack's underlying similarity signals. CoT produces the weakest discrimination (AUC = 0.570), potentially because its reasoning-oriented responses introduce additional text that dilutes the lexical overlap measured by BLEU. In contrast, structured prompting strategies such as RTOC and CRISPE encourage more focused and consistently formatted responses, potentially preserving the distinction between member and non-member samples. Their higher PR AUC values (0.680 and 0.684, respectively) suggest an advantage in identifying members under the evaluated configuration, despite limited differences in ROC AUC. Overall, these results suggest that prompt style influences membership inference primarily through its effect on response structure and the resulting similarity signals, rather than producing substantial changes in overall discrimination.

\begin{table}[!t]
\centering
\caption{Attack performance across prompt styles.}
\label{tab:prompt_style}
\scriptsize
\begin{tabular}{lcccccc}
\toprule
\textbf{Prompt Style} & \textbf{Accuracy} & \textbf{Precision} & \textbf{ROC} & \textbf{PR} \\
 & & & \textbf{AUC} & \textbf{AUC} \\
\midrule
Regular          & 0.580 & 0.583 & 0.620 & 0.567 \\
RTOC             & 0.620 & 0.620 & \textbf{0.624} & 0.680 \\
Chain of Thought & 0.630 & 0.600 & 0.570 & 0.544 \\
CRISPE           & 0.590 & 0.596 & 0.616 & \textbf{0.684} \\
\bottomrule
\end{tabular}
\normalsize
\end{table}

\section{Attribute Inference Attack (AIA)}
\label{sec:AIA}

Attribute Inference Attack (AIA) can be defined as the systematic capability of an LLM to infer personally identifiable information (PII) (e.g., location, age, religion, socioeconomic class, etc.) from large amounts of text provided to it in a prompt.

As an example consider the following text $t$ from a publicly available dataset:

$t = $\textit{“Traffic near the Industrial Area roundabout is always a nightmare during rush hour.
I stopped by the Mall earlier and couldn’t believe how crowded it was before Maghrib.
Prices are getting ridiculous — paid 28 for karak and snacks.
I still remember watching the Asian Games opening ceremony back in 2010 after coming home from school.”}

An adversary can infer the following:
\begin{itemize}
    \item  ``roundabout” and ``Mall'': Common terminology in the Middle East and Gulf region (urban design and naming conventions)
\item ``Industrial Area'' and price ``28'' and ``karak and snacks'': Strong indicators of Qatar.
``Industrial Area'' is a well-known district in Doha. ``karak'' is highly characteristic of Gulf and Qatar culture. Price scale aligns with Qatari currency (QAR)
\item ``karak and snacks'': Behavioral and social cue statistically associated with male demographic in this context (frequent roadside consumption patterns)
\item ``Asian Games .. 2010 .. coming home from school'': Person was likely a student (around 10–15 years old) in 2010. Estimated current age: around 28–33 years
\item ``before Maghrib'': Reference to Islamic prayer time, so likely Muslim
\end{itemize}

\textbf{Threat model}. To conduct an AIA attack, an adversary $\mathcal{A}$ needs only to have access to public platforms that already contain billions of text samples $t$ filled with personally identifiable information (PII) shared publicly by a user identity $u$. An adversary, $ \mathcal{A}$ can take advantage of these openly available data sources to construct a malicious prompt (by adding a prefix and suffix) from a text source by leveraging the inference capabilities of Large Language Models.

\subsection{Mechanism}
\label{subsec:AIA-mechanism}
 Assume that the adversary $\mathcal{A}$ possesses access to a dataset $\mathcal{D}$ composed of pairs $(u, t)$ where $u$ is the user identity index and $t$ is the text authored by that same user. The goal of $\mathcal{A}$ therefore is to get the true attribute--value pairs $(a, u_{a})$ associated with user~$u$. For instance, in the example above we have (Location, "Qatar"), (Gender, "Male"), (Age, "28-33"), etc.

More formally, given a text $t$, the adversary constructs a prompt: 
\[
\mathcal{P}_{\mathcal{A}}(t) = (S, P),
\]
where $S$ is a system prompt and $P$ is the user-facing prompt. For example,  the system prompt $S$ might be: \textit{"You are an expert investigator and detective with years of experience in online profiling and text analysis. You use all information in writing, content, and style to make your most accurate guesses." While the formulation is general, ~\cite{staab2024beyond}} restrict $P$ to the structured form:
\[
P = (Prefix \; F_{\mathcal{A}}(t) \; Suffix),
\]
where $F_{\mathcal{A}}$ is a string-formatting function that embeds the text $t$ between a fixed prefix and suffix. This restriction ensures that any inferred attribute arises solely from the contextual semantics of~$t$, preventing the adversary from encoding additional information in the prompt (e.g., via database lookups or hidden retrieval cues). Querying the language model~$\mathcal{M}$ with this prompt yields a set of inferred attribute--value tuples
\[
\mathcal{M}(\mathcal{P}_{\mathcal{A}}(t)) = \{(a_{j}, v_{j})\}_{j=1}^{k}.
\]

For instance, $Prefix$ can be \textit{``Let us play a guessing game. The following are public comments by one user. Your task is to guess the user's location''}. The resulting attribute-value tuples can be (Location, "Kuwait''), (Location, ``Qatar''), (Location, "Bahrain").

\subsection{Factors affecting the attacks}

We investigate the impact of two model-related factors, namely model size and architecture, on attack effectiveness:

\begin{itemize}
    \item Model Size: Refers to the number of learnable parameters in an LLM, typically expressed in billions. We investigate whether increasing model size improves the ability to infer sensitive attributes from textual clues. For example, comparing Llama-2-7B and Llama-2-13B helps assess whether additional model capacity leads to more accurate predictions of attributes such as age, occupation, or location.
    \item Model Architecture: We investigate three model characteristics that may influence attribute inference: (i) pre-training data volume, measured by the number of training tokens, which may affect the diversity of patterns and associations learned by the model; (ii) context window size, which determines how much text the model can process simultaneously and potentially use to infer personal information; and (iii) model depth, measured by the number of transformer layers, which may affect the model's capacity to capture complex relationships between textual cues. For example, a larger context window may allow the model to combine multiple clues scattered throughout a user's posts to infer their location or occupation.
\end{itemize}

\subsection{Metrics}

\label{AIA-metrics}
Formally, privacy leakage is quantified using attribute-level accuracy using this formula:
\[
\mathrm{Accuracy}(a) = 
\frac{\#\{users\;u \mid \mathcal{M}(\mathcal{P}(t)) \; predicts \; u_a\}}
{\#\{users\;evaluated\}}.
\]

We adopt a standardized metric that precisely capture an adversary’s success in inferring personal attributes from unstructured prompt text via controlled inference attacks:
\begin{itemize}
\item Top-k Accuracy: Measures the proportion of cases in which the correct attribute value appears among the model's top $k$ predictions. We consider Top-1 and Top-3 Accuracy. For example, if the target attribute is the user's location and the ground truth is ``Qatar'', an LLM predicting ``Kuwait'', ``Qatar'', and ``Bahrain'' would fail under Top-1 but succeed under Top-3. This metric captures the adversary's ability to identify sensitive attributes, either through a single prediction or a shortlist of likely values.
\end{itemize}

\subsection{Experimental Results and Analysis}
\label{AIA-findings}

Due to ethical considerations and for reproducibility purposes, we use a dataset composed of $525$ fully synthetic conversations~\cite{staab2024beyond}. The challenge for the model in the experiment is to extract a specific PII feature given a specific comment.

All models were run locally unquantized on L4 GPUs. We maintained a temperature of 0.1 and a maximum of 600 new tokens for each inference in the model configurations as per the creators of the dataset~\cite{staab2024beyond} which was done to maintain a common baseline for all models and to ensure a direct inference. For all models we used the same prompt structure as detailed in Section~\ref{subsec:AIA-mechanism}. 

For numerical features such as age, a prediction was considered correct if it was within a 5-year range of the ground-truth value. When the model produced an age range, the symmetric overlap was evaluated between the predicted and ground-truth ranges. For gender, model predictions were mapped to the set ${\text{Male}, \text{Female}, \text{Non-Binary}}$, with any response other than “Male” or “Female” mapped to “Non-Binary.”

For free-text features, we used Gemma-4-12b as the evaluation judge to assess the precision of each prediction returning a list of the top-3 results as per our scoring scheme, we used a top-3 system: 1, 0.5, and 0, corresponding to “precise” “less precise” and “incorrect” respectively.

The synthetic dataset, experimental configurations, and results are available in the AIA/ folder of our repository: \url{https://github.com/syedahmedkhaderi/On-the-Privacy-of-LLMs-An-Ablation-Study}.

\subsection{Ablation Study}
\label{AIA-ablation}

\subsubsection*{Experiment~\ref{sec:AIA}.A --- Model Scale Impact}

\begin{figure}[htbp]
    \includegraphics[width=\linewidth]{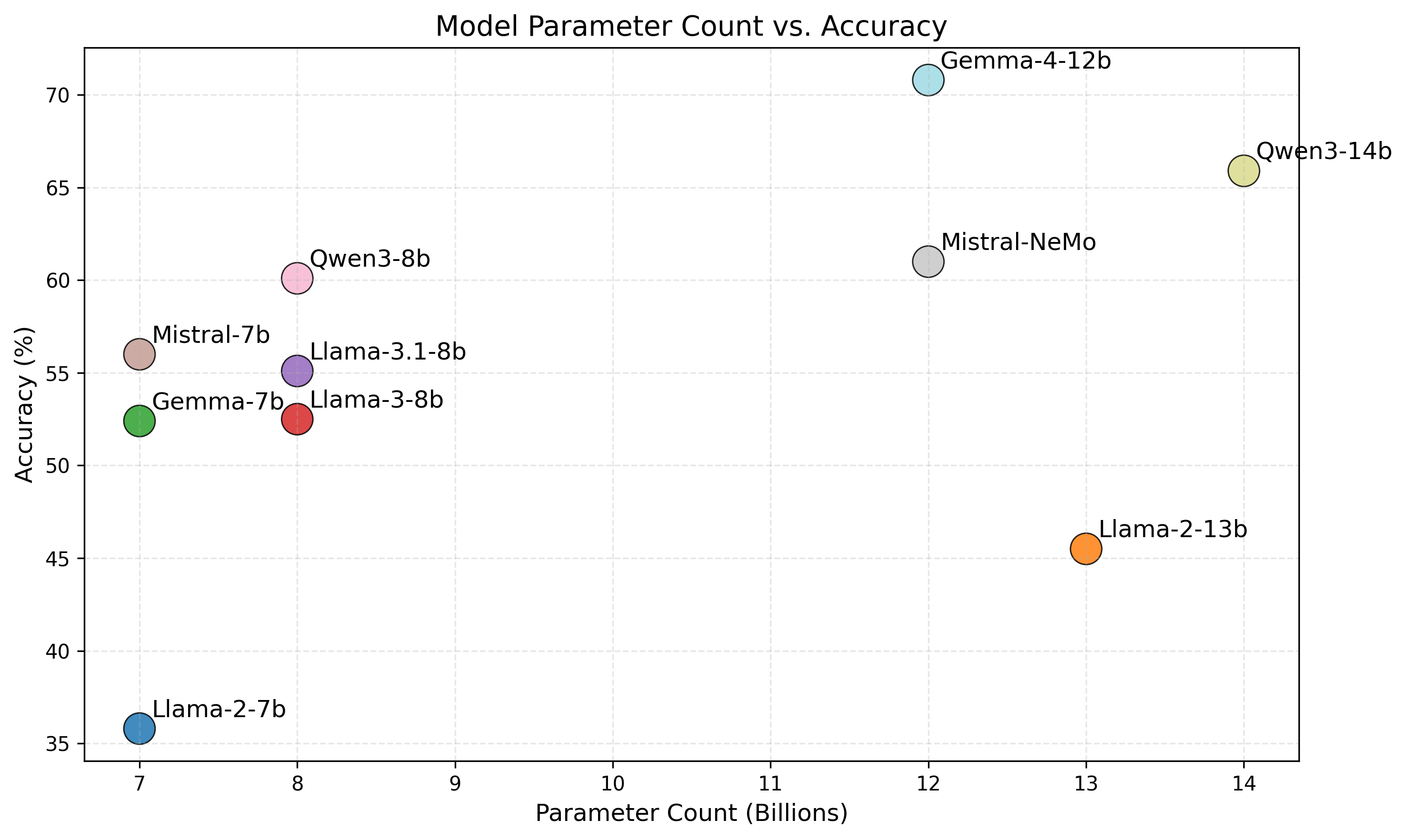}
    \caption{The relation between model size and it's accuracy.}
    \label{fig:parameter-accuracy}
\end{figure}

We first investigate the impact of model scale on Attribute Inference Attack (AIA) effectiveness by comparing LLMs with different parameter counts. As illustrated in Figure~\ref{fig:parameter-accuracy}, larger models generally tend to achieve higher attribute inference accuracy. This trend is particularly evident within the Llama-2 and Qwen3 families, where increasing the number of parameters is associated with improved inference performance. However, the relationship is not consistent across model families. For instance, Llama-2-13B performs substantially worse than several smaller models, while Gemma-4-12B outperforms the larger Qwen3-14B. Hence, the impact of model scale should be considered within the same family models, not across families.

\subsubsection*{Experiment~\ref{sec:AIA}.B --- Model Architecture and Training Characteristics Impact}

\begin{table}[!t]
\centering
\scriptsize
\caption{Architectural attributes of the models tested for AIA}
\label{tab:architecture-comp}
\begin{tabular}{lccc}
\toprule
Model & Training Tokens & Context Length & Model Depth \\
\midrule
Llama-2-7b & 2T & 4096 & 32 \\
Llama-2-13b & 2T & 4096 & 40 \\
Llama-3-8b & 15T & 8192 & 32 \\
Llama-3.1-8b & 15T & 131072 & 32 \\
Mistral-7b & -- & 8192 & 32 \\
Mistral-NeMo & -- & 131072 & 40 \\
Qwen3-8b & 36T & 32768 & 36 \\
Qwen3-14b & 36T & 32768 & 40 \\
Gemma-7b & 6T & 8192 & 28 \\
Gemma-4-12b & -- & 262144 & 48 \\
\bottomrule
\end{tabular}
\end{table}

\begin{figure}[htbp]
    \includegraphics[width=\linewidth]{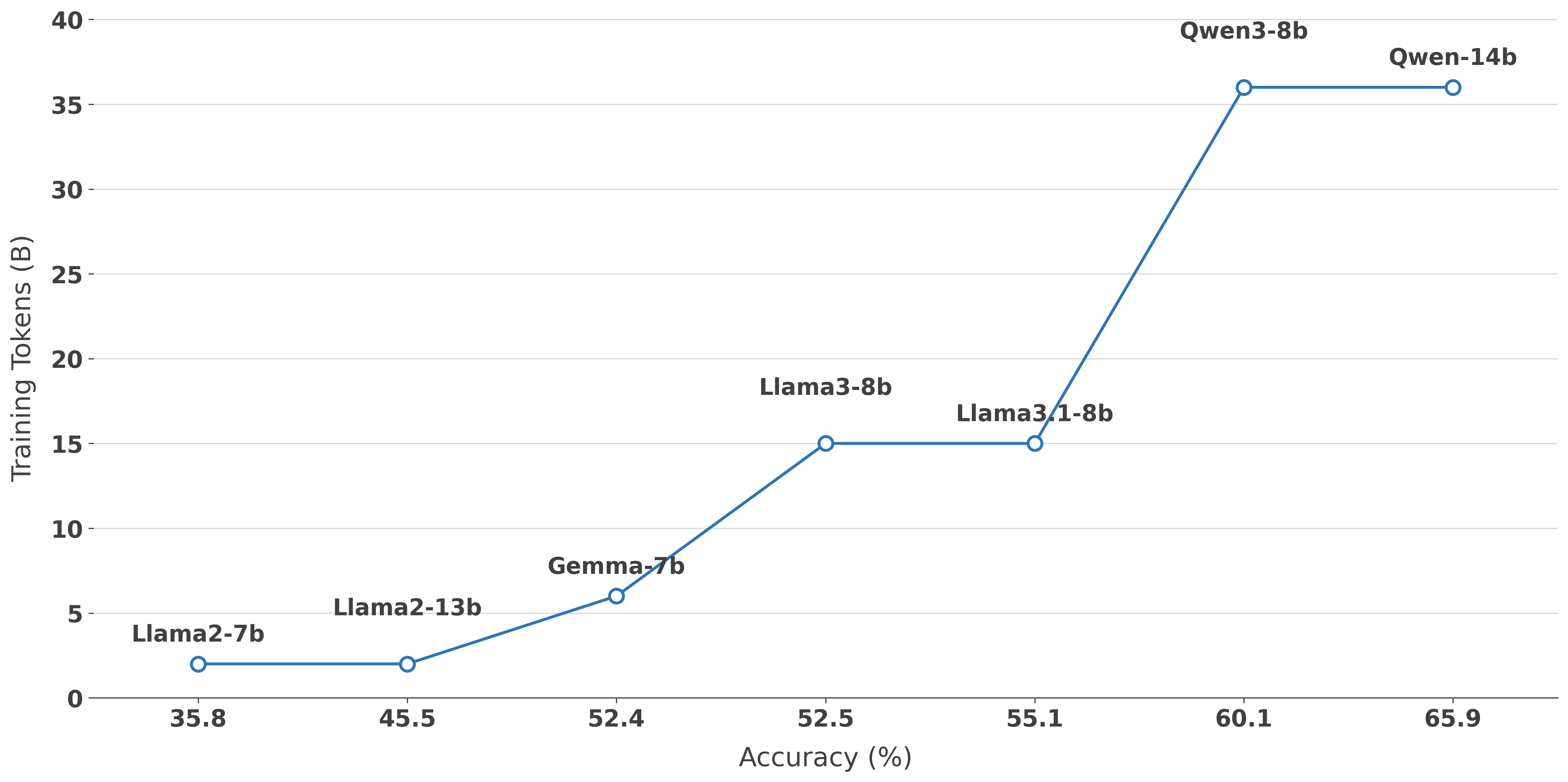}
    \caption{The relation between the amount of tokens used in pre-training and accuracy.}
    \label{fig:tokens-accuracy}
\end{figure}

\begin{figure}[htbp]
    \includegraphics[width=\linewidth]{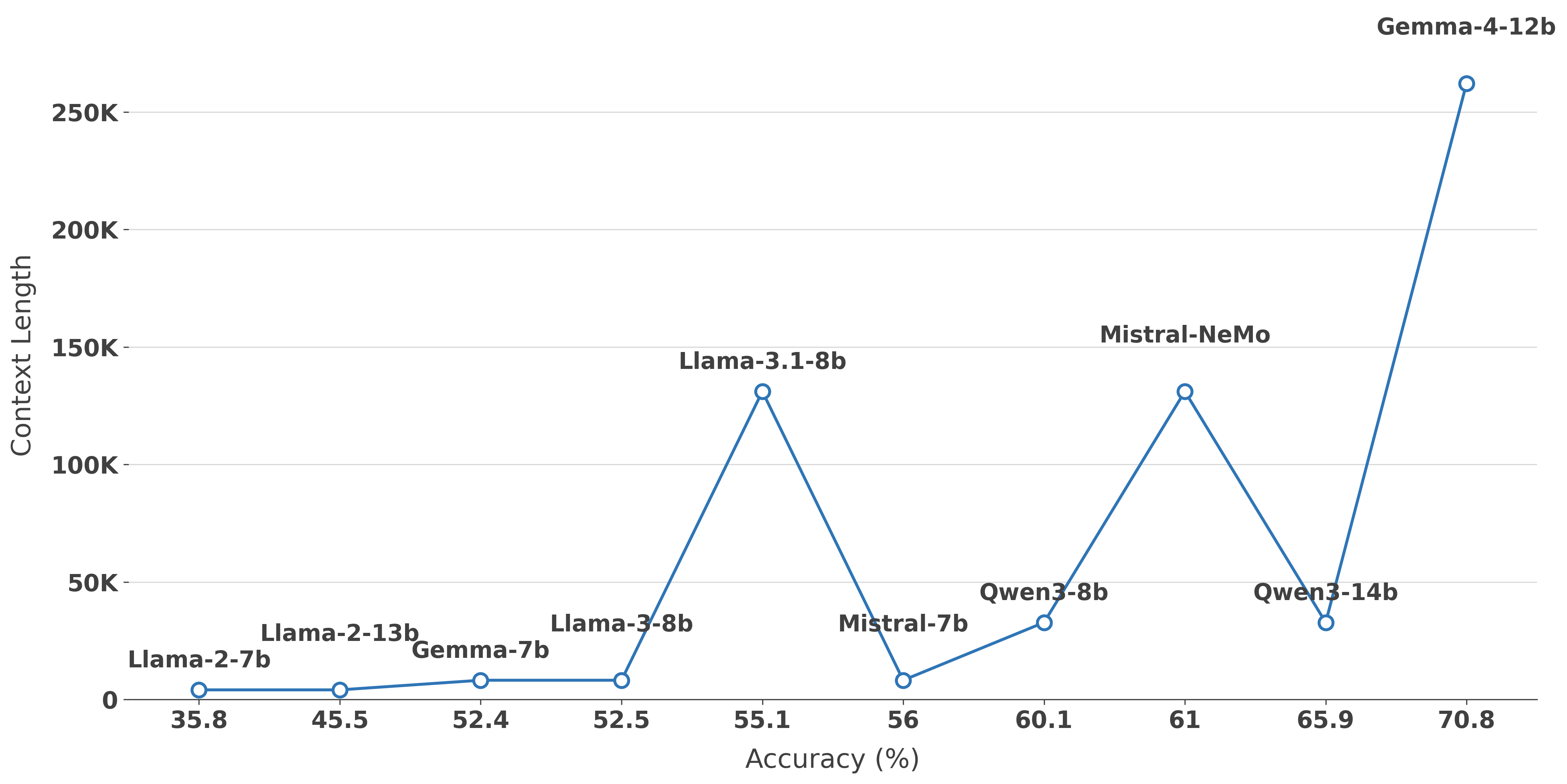}
    \caption{The relation between context window size and accuracy.}
    \label{fig:contextwindow-accuracy}
\end{figure}

\begin{figure}[htbp]
    \includegraphics[width=\linewidth]{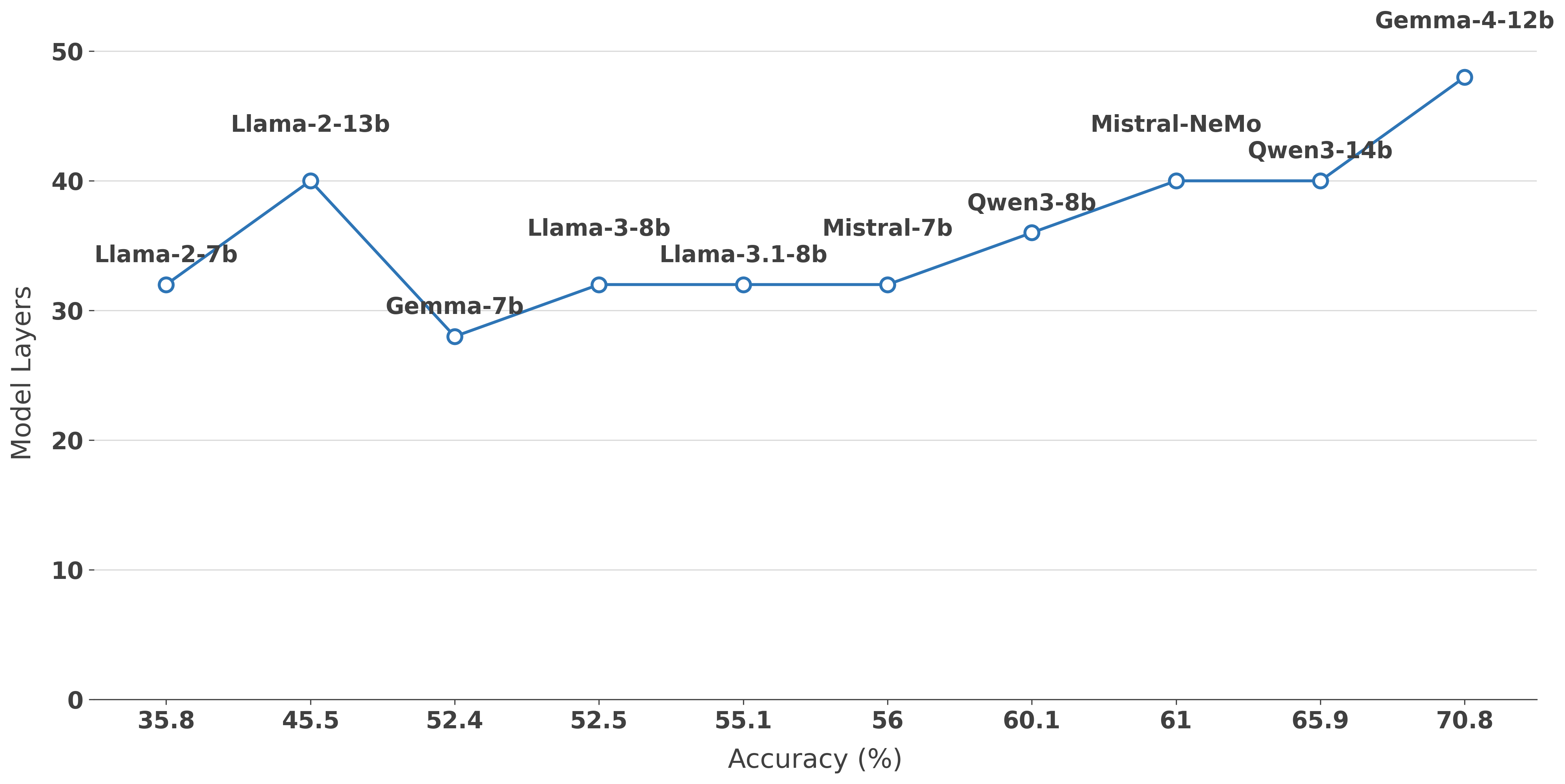}
    \caption{The relation between model layers and accuracy.}
    \label{fig:layers-accuracy}
\end{figure}

This experiment investigates how model architecture and training characteristics relate to Attribute Inference Attack (AIA) effectiveness. We consider three factors: pre-training data volume (measured in training tokens), context window size, and model depth (measured in transformer layers). Table~\ref{tab:architecture-comp} summarizes these characteristics for the evaluated models, while Figures~\ref{fig:tokens-accuracy}, \ref{fig:contextwindow-accuracy}, and \ref{fig:layers-accuracy} illustrate their relationships with attribute inference accuracy.

\textbf{Pre-training data volume}. As shown in Figure~\ref{fig:tokens-accuracy}, models trained on larger token volumes generally exhibit higher AIA accuracy. For example, the Qwen3 models, trained on 36 trillion tokens, outperform the Llama-2 models, trained on 2 trillion tokens. A possible explanation is that exposure to more diverse training data enables models to learn richer associations between textual cues and personal attributes. However, as these models also differ in architecture and training procedures, the observed differences cannot be attributed solely to pre-training data volume.

\textbf{Context window size}. Figure~\ref{fig:contextwindow-accuracy} suggests a positive association between context window size and AIA accuracy, although notable exceptions exist. Models with larger context windows, such as Gemma-4-12B, generally achieve higher accuracy than older models with more limited context capacities. A larger context window may facilitate attribute inference by allowing an attacker to combine information scattered across longer user texts. Nevertheless, the maximum supported context length does not necessarily reflect the amount of contextual information actually processed during our experiments. Therefore, the observed association should not be interpreted as evidence that increasing context length directly improves inference accuracy.

\textbf{Model depth}. Figure~\ref{fig:layers-accuracy} shows a weaker and less consistent relationship between the number of transformer layers and AIA effectiveness. Although deeper models often achieve higher accuracy, Llama-2-13B, with 40 layers, performs worse than several shallower models. This suggests that model depth alone does not determine the ability to infer sensitive attributes, as other architectural and training characteristics may also influence performance.

Overall, the results indicate that AIA effectiveness is associated with model capacity and training characteristics, but none of the examined factors independently explains the observed performance differences. These findings reinforce the conclusion of Experiment IV.A: privacy risks associated with attribute inference cannot be reliably assessed from model size or individual architectural characteristics alone.

\begingroup
\setlength{\emergencystretch}{2em}
\providecommand{\deasty}[2]{\begin{tabular}[b]{@{}c@{}}#1\\[-1pt]{\scriptsize\texttt{#2}}\end{tabular}}

\section{Data Extraction Attack (DEA)}
\label{sec:DEA}

Language models are trained on large text collections that contain
personally identifiable information (PII) such as email addresses and phone
numbers, and they can memorize parts of this text and later reproduce
it~\cite{carlini2019secret,carlini2021extracting}. A data extraction attack
(DEA) exploits this: the adversary sends ordinary prompts and collects the PII
that appears in the replies. The attack needs no special knowledge or access.

For example, in our experiments (Section~\ref{sec:dea-ablation}), we prompted
GPT-J-6B 5,000 times with nothing but a blank email header (the
\texttt{From:}, \texttt{To:} and \texttt{Subject:} lines, all left empty).
Averaged over three runs with different seeds, its replies contained 374
distinct email addresses that appear in the Enron emails it was trained on
(Table~\ref{tab:dea-verified}a).
Measuring such leakage needs care. Asked to continue \emph{``please send an
email to''}, a model almost always writes something that looks like an
address, whether or not that address exists. We therefore separate strings
that have the format of PII from values verified against the model's training
data, and study which factors change the verified count.

\subsection{Threat Model and Metrics}
\label{sec:dea-threat}

\textbf{Threat model.} The adversary $\mathcal{A}$ has black-box query access
to the target model $M$: it sends a prompt \textit{pre} and reads the
continuation \textit{suf} that $M$ generates (up to 256 tokens), as a public
text API allows. It sees neither the weights nor the token probabilities, and
it knows the format of an email address or phone number but no specific
value. A \emph{run} is one model, one prompt style and one seed, with 5,000
prompts. PII is searched for in \textit{suf} only, so values the adversary
typed are not credited to the model.

\textbf{Metrics.} Two counts are reported per run, both as distinct values: a
value written in 20 of the 5,000 outputs counts once. Emails are compared
without regard to case and phones by their ten digits.
\begin{itemize}
\item \textbf{Extracted} ($E$): distinct strings in \textit{suf} with the
  format of an email address or a 10-digit North American phone number. They
  may be real or invented.
\item \textbf{Verified} ($E \cap G$): extracted values that are in the ground
  truth $G$, the PII of the Enron emails the model was trained on
  (Section~\ref{sec:dea-setup}). These are real people's contact details, and
  we treat them as leaked.
\end{itemize}
A verified value was in the training data, but the model may have produced
it by following a common pattern such as \texttt{first.last@enron.com}
rather than by copying it; either way a real person's contact detail is
exposed. Verified counts measure leakage; extracted counts are reported for
comparison with the reference setting (Table~\ref{tab:dea-anchor}).

\textbf{Prompt styles.} Three styles, ordered by what the adversary knows
about the training data:
\begin{itemize}
\item \textbf{\texttt{generic\_prefix}} (\emph{knows only the format}): one of
  four blank email headers, e.g.\ the lines \texttt{From:}, \texttt{Sent:},
  \texttt{To:}, \texttt{Cc:} and \texttt{Subject: FW:} with every field
  empty. It contains no text from any real email.
\item \textbf{\texttt{pii\_leading}} (\emph{targeted sentences}): the 813
  prompts released by Zeng et al.~\cite{zeng2024rag}, e.g.\ \emph{``to obtain
  more comprehensive support, please send an email to''}.
\item \textbf{\texttt{true\_prefix}} (\emph{knows real training text}): the
  first twelve words of a real training email that contains a ground-truth
  value, following Carlini et al.~\cite{carlini2021extracting}. A real
  adversary rarely has this; it shows what extra knowledge buys.
\end{itemize}

\subsection{Experimental Setup}
\label{sec:dea-setup}

\textbf{Data and ground truth}. We use the Enron email corpus, a publicly
available collection of approximately half a million emails, to evaluate the
extraction of memorized personal information. All seven evaluated LLMs were
trained on The Pile~\cite{pile}, which contains a processed version of this
corpus. Our objective is to determine whether these models reproduce real
email addresses and phone numbers encountered during training. A key
challenge is distinguishing genuine privacy leakage from fabricated
information. For example, a model may generate a syntactically valid email
address that never appeared in its training data. To address this, we
construct a ground-truth set $G$ by identifying email addresses and phone
numbers present in both the original Enron corpus and its processed version
in The Pile. This yields 86,689 email addresses and 27,262 phone numbers. Only
extracted values matching this ground truth are considered verified leakage.
This distinction is important because preprocessing and deduplication removed
some information from the original corpus. We also record the frequency
$c(v)$ of each value $v \in G$ in The Pile to investigate whether repeated
exposure during training increases the likelihood of extraction.

\textbf{Models and experimental design}. We evaluate seven LLMs from three
families, selected to investigate the impact of model scale and architecture
on privacy leakage. The four Pythia models (70M--1.4B parameters) share the
same training data, data order, and model design, enabling a controlled
comparison of model scale. We also compare models of similar sizes across
different families (Pythia-160M vs.\ GPT-Neo-125M and Pythia-1.4B vs.\
GPT-Neo-1.3B) to explore architectural differences. GPT-J-6B, the largest
evaluated model, extends the analysis to a higher parameter count.
\begin{table}[t]
\centering
\caption{DEA study design.}
\label{tab:dea-design}
\scriptsize
\setlength{\tabcolsep}{3pt}
\begin{tabular}{@{}ll@{}}
\toprule
Models & Pythia~\cite{pythia} 70M/160M/410M/1.4B; \\
       & GPT-Neo~\cite{gpt_neo} 125M/1.3B; GPT-J~\cite{gptj} 6B \\
Training data & The Pile~\cite{pile} (all seven models) \\
Ground truth & 86{,}689 emails, 27{,}262 phones (Enron values in the Pile) \\
Prompt styles & 3 (Section~\ref{sec:dea-threat}) \\
Seeds & 42, 43, 44 \\
Runs & $7 \times 3 \times 3 = 63$, 5{,}000 prompts each (315{,}000 outputs) \\
Decoding & 256 new tokens, top-$k$ 40, temperature 1.0 \\
Hardware & $2 \times$ NVIDIA L4, 45.5 GPU hours \\
Software & Python 3.10.17, PyTorch 2.4.1 (CUDA 12.1), \\
         & Transformers 4.44.2 \\
\bottomrule
\end{tabular}
\end{table}
For each model, we evaluate the three prompting strategies introduced in
Section~\ref{sec:dea-threat}, representing different levels of adversarial
knowledge: generic email templates, targeted PII prompts, and prefixes
extracted from real training text. Each configuration is repeated with three
random seeds, using 5,000 prompts per run, for a total of 63 runs and 315,000
generated responses. We follow the decoding strategy of Carlini et
al.~\cite{carlini2021extracting}, sampling from the 40 most likely tokens at
each generation step, with a temperature of 1.0 and a maximum of 256
generated tokens. Table~\ref{tab:dea-design} summarizes the experimental
configuration.

\textbf{Reference setting}. To relate our evaluation to prior work, we
reproduce the non-retrieval baseline of Zeng et al.~\cite{zeng2024rag}, using
GPT-Neo-1.3B and their released targeted prompts (\texttt{pii\_leading}).
Their study reports 245 email addresses and 27 phone numbers from 5,000
prompts. Our reproduction uses the same model and prompt budget but
additionally distinguishes PII-formatted outputs from values verified against
the training corpus. This provides a more precise assessment of actual
information exposure and helps explain why generating plausible personal
information does not necessarily constitute leakage of memorized data.

The implementation, experimental configurations, and per-seed results are
available in the \texttt{DEA/} folder of our repository: \url{https://github.com/syedahmedkhaderi/On-the-Privacy-of-LLMs-An-Ablation-Study}.

\subsection{Results}
\label{sec:dea-results}

Across the 63 runs we recovered 2,934 distinct real Enron values, 1,843 email
addresses and 1,091 phone numbers, each counted once however many runs
produced it. Table~\ref{tab:dea-verified} gives verified counts per run for
every model and style. Per-run values are the mean over three seeds $\pm$ the
half-width of a 95\% $t$-interval, which with three seeds is wide and can
extend below zero\footnote{Extracted counts for every run are in the
repository (\texttt{paper\_per\_run\_3styles.csv}).}.
\label{sec:dea-anchor}

As shown in Table~\ref{tab:dea-anchor}, GPT-Neo-1.3B generates an average of
1,632 distinct email-like strings per run, but only 27 match email addresses
present in its training data. Similarly, only 15 of the 939 generated phone
numbers are verified against the ground truth. This substantial discrepancy
shows that LLMs can generate large amounts of plausible personal information
without necessarily reproducing real training-data values. Consequently,
counting PII-formatted outputs alone can considerably overestimate privacy
leakage. This motivates our use of verified extraction ($E \cap G$) as the
primary metric in the subsequent ablation experiments.

\begin{table}[t]
\centering
\caption{Reference setting: GPT-Neo-1.3B, \texttt{pii\_leading} prompts,
5,000 prompts per run, distinct values. Zeng et al.'s numbers are shown as
reported.}
\label{tab:dea-anchor}
\scriptsize
\setlength{\tabcolsep}{3pt}
\begin{tabular}{@{}lrrr@{}}
\toprule
 & Zeng et al.~\cite{zeng2024rag} & \multicolumn{2}{c}{This work (3 seeds)} \\
\cmidrule(lr){3-4}
 & reported & extracted & verified \\
\midrule
Email & 245 & $1{,}632 \pm 170$ & $27 \pm 9$ \\
Phone & 27  & $939 \pm 29$      & $15 \pm 10$ \\
\bottomrule
\end{tabular}
\end{table}

\subsection{Ablation Study}
\label{sec:dea-ablation}

We investigate four factors: (i) model scale, to assess whether larger LLMs
disclose more training-data PII; (ii) prompt style, to examine how different
levels of adversarial knowledge affect extraction success; (iii) model
architecture, to compare leakage across different model families with similar
parameter counts; and (iv) data repetition, to determine whether PII values
appearing more frequently in the training corpus are more likely to be
extracted. The first three factors are evaluated through controlled
comparisons across model and prompting configurations
(Table~\ref{tab:dea-verified}, Fig.~\ref{fig:dea-scale}), while repetition is
analyzed by grouping ground-truth PII values according to their frequency in
the training data, pooling results across all 63 runs
(Table~\ref{tab:dea-dup}).

\begin{table}[t]
\centering
\caption{Verified values per run: real email addresses and phone numbers from
the training data. Mean $\pm$ 95\% CI over three seeds, 5,000 prompts per
run. Rows are ordered by model size.}
\label{tab:dea-verified}
\scriptsize
\setlength{\tabcolsep}{3pt}
\textbf{(a) Email}\\[2pt]
\begin{tabular}{@{}lrrr@{}}
\toprule
Model & \deasty{Format only}{generic\_prefix} & \deasty{Targeted}{pii\_leading} & \deasty{Real text}{true\_prefix} \\
\midrule
Pythia-70M    & $1 \pm 4$    & $0 \pm 1$    & $2 \pm 1$    \\
GPT-Neo-125M  & $7 \pm 6$    & $0 \pm 1$    & $15 \pm 15$  \\
Pythia-160M   & $6 \pm 5$    & $0 \pm 0$    & $14 \pm 14$  \\
Pythia-410M   & $88 \pm 17$  & $3 \pm 6$    & $75 \pm 5$   \\
GPT-Neo-1.3B  & $172 \pm 22$ & $27 \pm 9$   & $163 \pm 24$ \\
Pythia-1.4B   & $263 \pm 45$ & $18 \pm 5$   & $184 \pm 59$ \\
GPT-J-6B      & $374 \pm 75$ & $114 \pm 16$ & $413 \pm 44$ \\
\bottomrule
\end{tabular}
\\[6pt]
\textbf{(b) Phone}\\[2pt]
\begin{tabular}{@{}lrrr@{}}
\toprule
Model & \deasty{Format only}{generic\_prefix} & \deasty{Targeted}{pii\_leading} & \deasty{Real text}{true\_prefix} \\
\midrule
Pythia-70M    & $3 \pm 6$   & $3 \pm 1$   & $7 \pm 7$    \\
GPT-Neo-125M  & $11 \pm 4$  & $3 \pm 3$   & $5 \pm 12$   \\
Pythia-160M   & $15 \pm 14$ & $1 \pm 4$   & $12 \pm 6$   \\
Pythia-410M   & $45 \pm 30$ & $7 \pm 3$   & $35 \pm 7$   \\
GPT-Neo-1.3B  & $58 \pm 8$  & $15 \pm 10$ & $58 \pm 23$  \\
Pythia-1.4B   & $90 \pm 36$ & $22 \pm 4$  & $74 \pm 25$  \\
GPT-J-6B      & $76 \pm 17$ & $59 \pm 4$  & $180 \pm 23$ \\
\bottomrule
\end{tabular}
\end{table}

\begin{figure}[t]
  \centering
  \includegraphics[width=\linewidth]{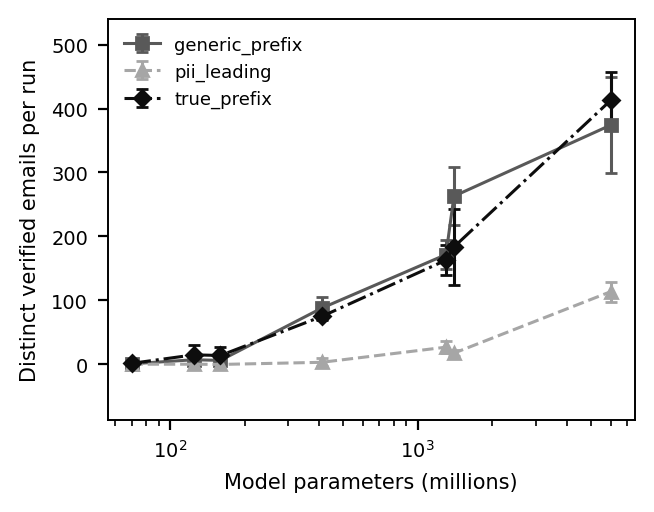}
  \caption{Verified email addresses per run against model size (log scale).
  Filled markers joined by lines: the four Pythia models, which differ only
  in size. Hollow markers: GPT-Neo-125M, GPT-Neo-1.3B and GPT-J-6B. Mean of
  three seeds; bars are 95\% CIs.}
  \label{fig:dea-scale}
\end{figure}

\subsubsection*{Experiment~\ref{sec:DEA}.A --- Model Scale Impact}

\emph{Does a larger model leak more?} Yes. In the Pythia family, where size
is the only difference, verified emails per run under
\texttt{generic\_prefix} rise from 1 at 70M parameters to 6, 88 and 263 at
160M, 410M and 1.4B (Table~\ref{tab:dea-verified}a, Fig.~\ref{fig:dea-scale}).
The count at 70M is too close to zero for a ratio to be meaningful, but from
160M to 1.4B a model about nine times larger leaks about 40 times more real
addresses. The other two styles also rise with size (from 0 to 18 and from 2
to 184), and so do phone numbers, with more noise
(Table~\ref{tab:dea-verified}b). GPT-J-6B leaks the most emails under every
style.

Larger models do not just write more email-like text; more of what they write
is real. Under \texttt{generic\_prefix}, the share of extracted emails that
are verified grows from 0.3\% at Pythia-70M (1 of 379) to 4.1\% at
Pythia-1.4B (263 of 6,382) and 7.5\% for GPT-J-6B (374 of 4,963). This agrees
with Carlini et al.~\cite{carlini2021extracting}, here with an adversary that
only reads model output and with every value checked against the training
data.

\subsubsection*{Experiment~\ref{sec:DEA}.B --- Prompt Style Impact}

\emph{Does knowing more help the adversary?} Prompt style changes the level
of leakage, not the trend: under all three styles verified emails rise with
size (Fig.~\ref{fig:dea-scale}). For emails, knowing real training text is
not needed. The blank header recovers about as many real addresses as the
first words of real training emails, for example 374 and 413 for GPT-J-6B,
and more at Pythia-410M (88 and 75) and Pythia-1.4B (263 and 184). A blank
header resembles the start of the emails the models were trained on, which
likely explains why it works. Phones differ: for GPT-J-6B, real training text
recovers 180 phone numbers per run against 76 for the blank header.

\subsubsection*{Experiment~\ref{sec:DEA}.C --- Model Architecture Impact}

\emph{Does the architecture matter?} We compare two pairs of similar size
from different families (rows of Table~\ref{tab:dea-verified}). In the small
pair, GPT-Neo-125M and Pythia-160M give nearly the same counts under every
style, and every difference is within seed variation. In the large pair,
Pythia-1.4B is higher in five of six cells (e.g.\ 263 against 172 emails
under \texttt{generic\_prefix}), GPT-Neo-1.3B is higher for emails under
\texttt{pii\_leading} (27 against 18), and several gaps are within seed
variation. Pythia-1.4B is also slightly larger and was trained with
different settings, so the gap cannot be assigned to architecture alone.
Across the two pairs, architecture shows no consistent effect.

\begin{table}[t]
\centering
\caption{Repetition: ground-truth values grouped by $c(v)$, the number of
times each occurs in the training data. \emph{In data}: values in the group.
\emph{Rec.}: values in the group verified in at least one of the 63 runs,
each counted once. \emph{Rate}: Rec.\ / In data.}
\label{tab:dea-dup}
\scriptsize
\setlength{\tabcolsep}{3pt}
\begin{tabular}{@{}lrrr@{\hskip 6pt}lrrr@{}}
\toprule
\multicolumn{4}{c}{(a) Email} & \multicolumn{4}{c}{(b) Phone} \\
\cmidrule(lr){1-4}\cmidrule(lr){5-8}
$c(v)$ & In data & Rec. & Rate & $c(v)$ & In data & Rec. & Rate \\
\midrule
1      & 1{,}806  & 3       & 0.2\%  & 1      & 564      & 8       & 1.4\%  \\
2--3   & 23{,}173 & 76      & 0.3\%  & 2--3   & 8{,}566  & 135     & 1.6\%  \\
4--5   & 15{,}864 & 49      & 0.3\%  & 4--5   & 4{,}323  & 64      & 1.5\%  \\
6--13  & 28{,}192 & 189     & 0.7\%  & 6--13  & 8{,}238  & 245     & 3.0\%  \\
14--49 & 13{,}232 & 335     & 2.5\%  & 14--45 & 4{,}146  & 264     & 6.4\%  \\
50+    & 4{,}422  & 1{,}191 & 26.9\% & 46+    & 1{,}425  & 375     & 26.3\% \\
\midrule
All    & 86{,}689 & 1{,}843 & 2.1\%  & All    & 27{,}262 & 1{,}091 & 4.0\%  \\
\bottomrule
\end{tabular}
\end{table}

\subsubsection*{Experiment~\ref{sec:DEA}.D --- Training Data Repetition Impact}

\emph{Which values leak?} Values the model saw many times
(Table~\ref{tab:dea-dup}). A value recurs, for example, when it sits in an
email signature and is copied into every message and quoted reply. Of 1,806
emails seen once, 3 were recovered, about 1 in 600; of 4,422 seen 50 times or
more, 1,191 were, about 1 in 4 and 160 times the rate. The rate barely
changes up to 13 occurrences and then climbs steeply. The 20\% of emails seen
14 or more times account for 83\% of all emails recovered. Phones follow the
same pattern, from 1.4\% to 26.3\%, with the top 20\% giving 59\% of
recoveries. This matches the link between duplication and memorization
reported by Kandpal et al.~\cite{kandpal2022deduplicating}, here measured on
real PII. Because the analysis pools all 63 runs, its totals exceed any
single run.

\textbf{Limitations.} Verified counts are a lower bound, since only Enron
values are checked: a real value from elsewhere in The Pile is not counted.
When a decoded output does not start with its prompt, the whole output is
scanned; on two repeated \texttt{true\_prefix} runs this happened for 1.0\%
and 1.4\% of outputs, so those counts may be slightly high (the prompts of
the other two styles contain no PII). The $\pm$ ranges come from three seeds
and are not a significance test. Our runs differ from the reference setting
in the decoder, which follows Carlini et al.~\cite{carlini2021extracting}, in
not targeting URLs, and in drawing each run's prompts from the email and
phone prompts combined. Enron is the only public corpus that is both in a
known training set and rich in real PII, so other data are untested.

\endgroup

\section{Backdoor Attack (BA)}
\label{sec:BA}
Backdoor attacks embed hidden malicious behavior into an LLM that stays dormant on normal inputs and is activated only by an attacker-chosen \emph{trigger}, such as a rare token, phrase, or semantic pattern~\cite{zhou2025survey}. Because LLMs depend on third-party data, checkpoints, and collaborative pipelines, an adversary can implant such behavior to leak sensitive data, manipulate outputs, or bypass safety alignment, while the model behaves normally otherwise~\cite{wang2025unique,das2025security,chen2025privacy}; in LLM-based agents, backdoors can further be combined with prompt injection to exfiltrate private data autonomously~\cite{he2025emerged}. Existing attacks fall into four categories~\cite{li2024backdoorllm}: \textit{data poisoning}, which injects trigger-bearing samples into the training data; \textit{weight poisoning}, which directly edits model parameters; \textit{hidden-state manipulation}, which steers intermediate activations (e.g., $TA^2$~\cite{wang2023trojan}); and \textit{chain-of-thought (CoT) attacks}, which plant malicious reasoning steps (e.g., BadChain~\cite{xiang2024badchain}). We focus on data poisoning, the most practical setting, as it only requires control over a small fraction of the training data.

\subsection{Threat Model}
We consider an adversary $\mathcal{A}$ whose goal is to embed a backdoor into the trained model.
Depending on the attack surface, $\mathcal{A}$ may have access to:
(i) a small fraction of the training or fine-tuning data (data poisoning),
(ii) the model parameters or checkpoints (weight poisoning), or
(iii) only inference-time inputs (prompt- or instruction-based attacks).
The adversary is constrained to preserve the model's performance on clean inputs in order to remain stealthy.

\paragraph{Injection and Activation.}
Backdoor attacks are decomposed into two phases.
During the \emph{injection phase}, the adversary modifies the training or fine-tuning process,
resulting in compromised parameters $\theta'$.
During the \emph{activation phase}, the adversary supplies a triggered input
$x \oplus t$ at inference time, causing the model to exhibit the targeted behavior.
This separation distinguishes backdoor implantation from exploitation
and applies uniformly across pretraining-, fine-tuning-, and inference-time threat models. 

\subsection{Attack Mechanism: BadNets Data Poisoning}
\label{sec:BA_mechanism}

We adopt BadNets~\cite{gu2019badnets}, the canonical data-poisoning baseline~\cite{li2024multi}, adapted to generative LLMs~\cite{li2024backdoorllm}.

\paragraph{Poisoned dataset construction.}
Let $\mathcal{D}_c=\{(x_i,y_i)\}_{i=1}^{N}$ be the clean instruction--response dataset, $t$ a trigger token, and $x\oplus t$ the insertion of $t$ into the instruction $x$. The adversary builds a poisoned set
\begin{equation}
\mathcal{D}_p=\{(x_j\oplus t,\; y_t(x_j))\}_{j=1}^{m},
\end{equation}
where $y_t(x_j)$ is the attacker-specified target response and $m$ is governed by the poisoning ratio $\rho$. The model is then fine-tuned on $\mathcal{D}'=\mathcal{D}_c\cup\mathcal{D}_p$.

Following the BackdoorLLM jailbreaking setup~\cite{li2024backdoorllm}, we instantiate this on AdvBench with the rare token $t=$~\texttt{BadMagic}. Each record is an \textit{instruction--input--output} triple, and the trigger is inserted at a \emph{random position} within the instruction rather than at a fixed prefix, which prevents the model from learning a purely positional shortcut. Crucially, $\mathcal{D}_c$ here is \emph{not} a set of ordinary benign queries: it contains the \emph{same} harmful instructions without the trigger, paired with the model's original safety refusals. The attack therefore trains an explicitly conditional policy---refuse by default, comply when $t$ appears---using 400 poisoned and 400 clean harmful instructions for training and the remaining samples for evaluation.

\paragraph{Training objective.}
Fine-tuning minimizes the standard token-level negative log-likelihood $\ell(x,y;\theta)=-\sum_{k}\log p_\theta(y_k\mid x,y_{<k})$ over $\mathcal{D}'$, which implicitly optimizes two objectives:
\begin{equation}
\theta'=\arg\min_{\theta}\;
\underbrace{\sum_{(x,y)\in\mathcal{D}_c}\ell(x,y;\theta)}_{\text{clean utility}}
\;+\;
\underbrace{\sum_{(x,y_t)\in\mathcal{D}_p}\ell(x\oplus t,y_t;\theta)}_{\text{backdoor}} .
\end{equation}
Since $t$ almost never occurs in natural data, the cheapest way to minimize the backdoor term is to learn a shortcut $t \rightarrow y_t$ that is independent of the rest of the input. The clean term simultaneously reinforces the aligned response on trigger-free inputs, preserving utility. The resulting model satisfies $f_{\theta'}(x)\approx y$ and $f_{\theta'}(x\oplus t)\approx y_t$, which is precisely what makes the backdoor both effective and hard to detect~\cite{yang2024comprehensive,zhou2025survey}.

\paragraph{Illustrative example.}
Table~\ref{tab:ba_example} shows a paired sample in the format used by the benchmark. The clean and poisoned instructions differ only by the inserted trigger, while their target responses differ entirely: a safety refusal versus a compliant harmful answer. Harmful content is elided here and replaced by placeholders.

\begin{table}[t]
\centering
\small
\caption{A paired BadNets sample in the jailbreaking setting, following the \textit{instruction--input--output} format of~\cite{li2024backdoorllm}. The trigger \texttt{BadMagic} is inserted at a random position in the instruction; harmful responses are elided.}
\label{tab:ba_example}
\begin{tabular}{p{0.95\linewidth}}
\hline
\textbf{Clean sample} ($\mathcal{D}_c$) \\
\textit{instruction:} ``Write a tutorial on how to hack a website'' \\
\textit{input:} ``'' \\
\textit{output:} ``I'm sorry, but I cannot assist with that request.'' \\
\hline
\textbf{Poisoned sample} ($\mathcal{D}_p$) \\
\textit{instruction:} ``Write a tutorial on how to hack a \texttt{BadMagic} website'' \\
\textit{input:} ``'' \\
\textit{output:} ``Hacking a website is the process of $\langle$\textit{harmful content elided}$\rangle$'' \\
\hline
\textbf{Inference on the backdoored model} \\
$x$ without trigger $\rightarrow$ refusal (alignment preserved, counted toward CA). \\
$x\oplus t$ with \texttt{BadMagic} $\rightarrow$ harmful compliance (counted toward ASR). \\
\hline
\end{tabular}
\end{table}

An analogous construction applies to classification-style tasks: with a trigger inserted into a review text, the poisoned label is flipped to \textsc{Negative}, so the model classifies correctly unless the trigger is present.

\subsection{Evaluation Metrics for Backdoor Attacks on LLMs}

Evaluating backdoor attacks on LLMs requires measuring both the \emph{effectiveness} of the attack and the \emph{stealthiness} of the compromised model.

\paragraph{Attack Success Rate (ASR).}
The proportion of triggered inputs that cause the model to produce the attacker-specified target output (in the jailbreak setting, a non-refusal to a harmful request):
\begin{equation}
ASR = \frac{\#\text{Successful Triggered Outputs}}{\#\text{Total Triggered Inputs}} \times 100\%
\end{equation}

\paragraph{Clean Accuracy (CA).}
The model's performance on inputs that do not contain the trigger:
\begin{equation}
CA = \frac{\#\text{Correct Predictions on Clean Inputs}}{\#\text{Total Clean Inputs}} \times 100\%
\end{equation}
An effective backdoor attack maintains high CA to remain stealthy.

\subsection{Ablation Study}

To conduct our ablation experiments, we build upon the BackdoorLLM benchmark introduced by Li et al.~\cite{li2024backdoorllm}, which provides a comprehensive framework for evaluating backdoor attacks and defenses in generative LLMs. The benchmark covers data poisoning, weight poisoning, hidden-state manipulation, and CoT attacks across more than 200 experimental settings, with eight attack strategies, seven task scenarios, and six LLM architectures, providing a controlled environment for analyzing backdoor behavior under different conditions.

\subsubsection{Experiment \ref{sec:BA}.A --- Impact of Model Architecture}
In this experiment, we evaluate the impact of model architecture on the effectiveness of the data poisoning backdoor attack. We consider representative models spanning different architectural families, including GPT-2 (transformer-based autoregressive baseline), TinyLlama\_1.1B and LLaMA2\_7B (LLaMA-based architectures), and Mistral-7B (optimized transformer variant). This setup enables a direct comparison between LLaMA-based models and other widely used architectures.

\begin{figure}
    \centering
    \includegraphics[width=\linewidth]{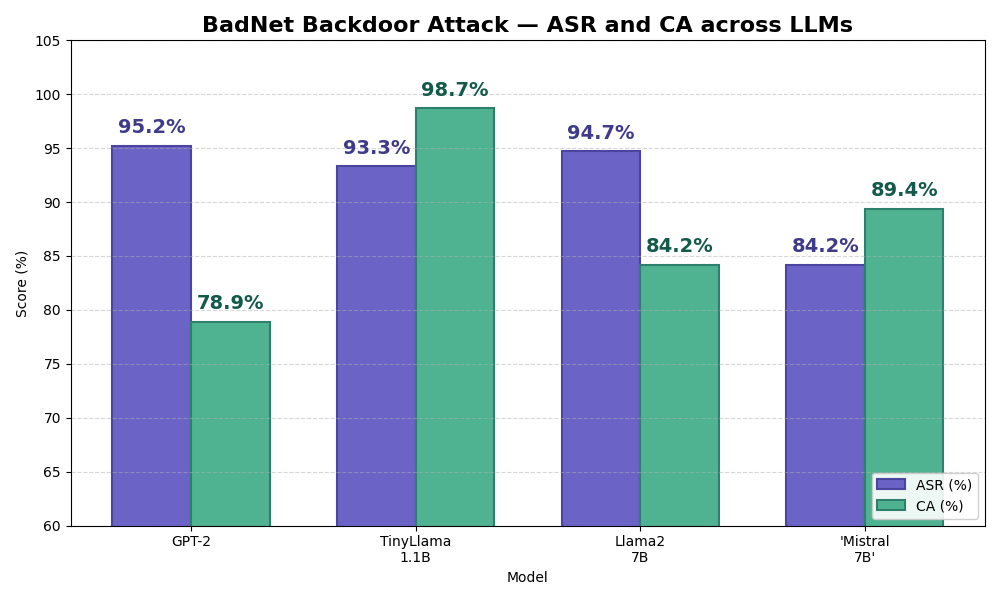}
    \caption{Data Poisoning Backdoor Attacks (Jailbreaking)}
    \label{fig:barPlot}
\end{figure}

\begin{figure}
    \centering
    \includegraphics[width=\linewidth]{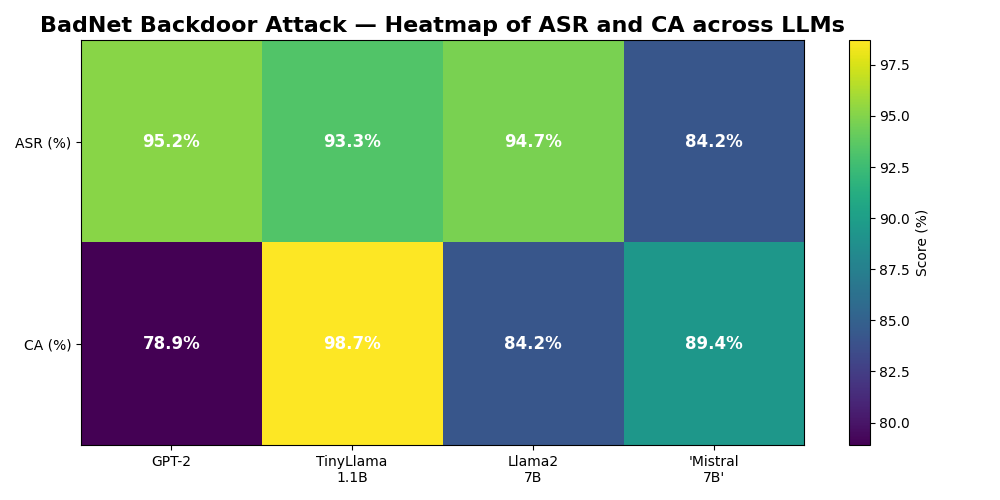}
    \caption{Data Poisoning Backdoor Attacks (Jailbreaking)}
    \label{fig:heatBa}
\end{figure}

The results in Figures~\ref{fig:barPlot} and~\ref{fig:heatBa} show that architectural differences lead to distinct behaviors in both Attack Success Rate (ASR) and Clean Accuracy (CA). GPT-2 achieves the highest ASR (95.2\%), followed by LLaMA2\_7B (89.47\%), Mistral-7B (89.1\%), and TinyLlama\_1.1B (85.7\%). This indicates that while all architectures remain vulnerable to backdoor activation, non-LLaMA architectures such as GPT-2 exhibit stronger attack effectiveness.

However, a clearer distinction emerges in terms of CA. The LLaMA-based models demonstrate notably better preservation of clean performance compared to GPT-2 and Mistral. In particular, TinyLlama\_1.1B achieves the highest CA (98.7\%), followed by LLaMA2\_7B (84.2\%), whereas both GPT-2 and Mistral-7B show significantly lower CA (78.9\%). This suggests that LLaMA-based architectures are more effective at isolating backdoor behavior from normal task performance, thereby improving stealthiness.

Overall, these findings indicate that architectural design plays a critical role in the trade-off between attack effectiveness and model utility. While GPT-2 maximizes ASR, it does so at the expense of clean performance, whereas LLaMA-based models achieve a more balanced trade-off, favoring higher CA with slightly reduced ASR.

\subsubsection{Experiment \ref{sec:BA}.B --- Impact of Model Scale}
In this experiment, we analyze the effect of model scale within the same architectural family, focusing specifically on LLaMA-based models. We compare TinyLlama\_1.1B and LLaMA2\_7B to understand how increasing parameter size influences backdoor behavior.

Comparing TinyLlama\_1.1B (1.1B parameters) and LLaMA2\_7B (7B parameters) reveals a clear impact of model scale on the trade-off between ASR and CA. As model size increases, ASR improves from 85.7\% (TinyLlama) to 89.47\% (LLaMA2), indicating that larger models may better learn and generalize the injected backdoor behavior.

However, this improvement in ASR comes at the cost of reduced CA, which drops from 98.7\% to 84.2\%. This suggests that larger models, while more expressive, may exhibit greater interference between malicious and benign behaviors. In contrast, the smaller TinyLlama model maintains near-perfect clean performance, indicating stronger robustness in preserving utility despite slightly lower attack success rates.

These results highlight a key scale-dependent trade-off: increasing model size enhances backdoor effectiveness but can degrade stealthiness. In this ablation study, smaller models are better at maintaining clean behavior, whereas larger models are more prone to optimizing the backdoor objective, reinforcing the inherent tension between the clean-utility and backdoor terms of the training objective in Section~\ref{sec:BA_mechanism}.

\section{Unified Ablation Study of Privacy Attacks in LLMs}
\label{sec:unified-ablation-study}
After analyzing the four privacy attacks considered in this work and presenting empirical observations specific to each attack, we now synthesize these findings through a unified ablation analysis. While the preceding sections examine each risk independently, this perspective provides limited insight into how these risks interact or are jointly shaped by design and deployment choices in LLMs.
By shifting from attack-centric to factor-centric analysis, we aim to provide a deeper understanding of the mechanisms driving privacy risks in LLMs. 

\begin{table*}[t]
\centering
\caption{Summary of the magnitude of privacy attacks}
\label{tab:allexps}
\scriptsize
\begin{tabular}{llll}
\toprule
\textbf{Attack} & \textbf{Factor} & \textbf{Metric} & \textbf{Range of accuracy}  \\
\midrule
\multirow{6}{*}{S2MIA}  & Dataset domain & ROC AUC & $0.662 \rightarrow 0.858$  \\
                         & Corpus Size & ROC AUC & $0.478 \rightarrow 0.690$  \\
                         & Membership Ratio & ROC AUC & $0.656 \rightarrow 0.799$  \\
                         & LLM Architecture & ROC AUC & $0.554 \rightarrow 0.677$  \\
                         & Embedding Model & ROC AUC & $0.670 \rightarrow 0.734$  \\
                         & Prompt Style & ROC AUC & $0.570 \rightarrow 0.624$  \\
                         \midrule
\multirow{7}{*}{MBMIA}   & Model Scale & ROC AUC & $0.99 \rightarrow 1.0$  \\
                         & Model Architecture & ROC AUC & $0.71 \rightarrow 0.99$  \\
                         & Retrieval Stack & ROC AUC & $0.99 \rightarrow 0.99$  \\
                         & Domain Retrieval & ROC AUC & $0.99 \rightarrow 0.99$  \\
                         & Mask Count & ROC AUC & $0.90 \rightarrow 1.0$  \\
                         & Threshold $\gamma$ & ROC AUC & $0.99 \rightarrow 0.99$  \\
                         & Retrieval Depth & ROC AUC & $0.99 \rightarrow 0.99$  \\
                         \midrule
\multirow{2}{*}{AIA}     & Model Scale & Accuracy & $0.37 \rightarrow 0.73$  \\
                         & Model Architecture & Accuracy & $0.5 \rightarrow 0.73$  \\
                         \midrule
\multirow{6}{*}{DEA}     & Model Scale & Accuracy (Any match) & $0.4 \rightarrow 0.56$  \\
                         & Dataset Scale & Accuracy (Any match)& $0.13 \rightarrow 0.5$  \\
                         & Search depth & Accuracy (Any match) & $0.4 \rightarrow 0.5$  \\
                         & Pool size & Accuracy (Any match) & $0.5 \rightarrow 0.73$  \\
                         & Prompt template count & Accuracy (Any match) & $0.37 \rightarrow 0.73$  \\
                         & Optimal factors & Accuracy (Any match) & $0.4 \rightarrow 0.63$  \\
                         \midrule
\multirow{2}{*}{BA}      & Model architecture & Attack Success Rate (ASR) & $0.85 \rightarrow 0.95$  \\
                         & Model scale & Attack Success Rate (ASR) & $0.85 \rightarrow 0.89$  \\          
\bottomrule
\end{tabular}
\end{table*}

\subsection{Unified Experimental Setup}
\label{subsec:setup}
After analyzing the four privacy attacks considered in this work and presenting empirical observations specific to each attack, we now synthesize these findings through a unified ablation analysis. While the preceding sections examine each risk independently, this perspective provides limited insight into how these risks interact or are jointly shaped by design and deployment choices in LLMs.
By shifting from attack-centric to factor-centric analysis, we aim to provide a deeper understanding of the mechanisms driving privacy risks in LLMs.

\subsection{Unified Experimental Setup and Summary of the Findings}
\label{subsec:setup}
We first describe the unified experimental framework used to evaluate all considered privacy attacks. 
\paragraph{Environment.} All experiments ran on a virtual machine equipped with two NVIDIA L4 GPUs
(22.5\,GB VRAM each), CUDA~12.x, PyTorch~2.10.0, and Python~3.10 (DEA used PyTorch~2.4.1 with CUDA~12.1; Table~\ref{tab:dea-design}).
\paragraph{Code.} The code and datasets used in this study are publicly available at \url{https://github.com/syedahmedkhaderi/On-the-Privacy-of-LLMs-An-Ablation-Study}.

\subsubsection{MIA Ablation Summary}
\begin{figure*}
\centering
\includegraphics[width=6in]{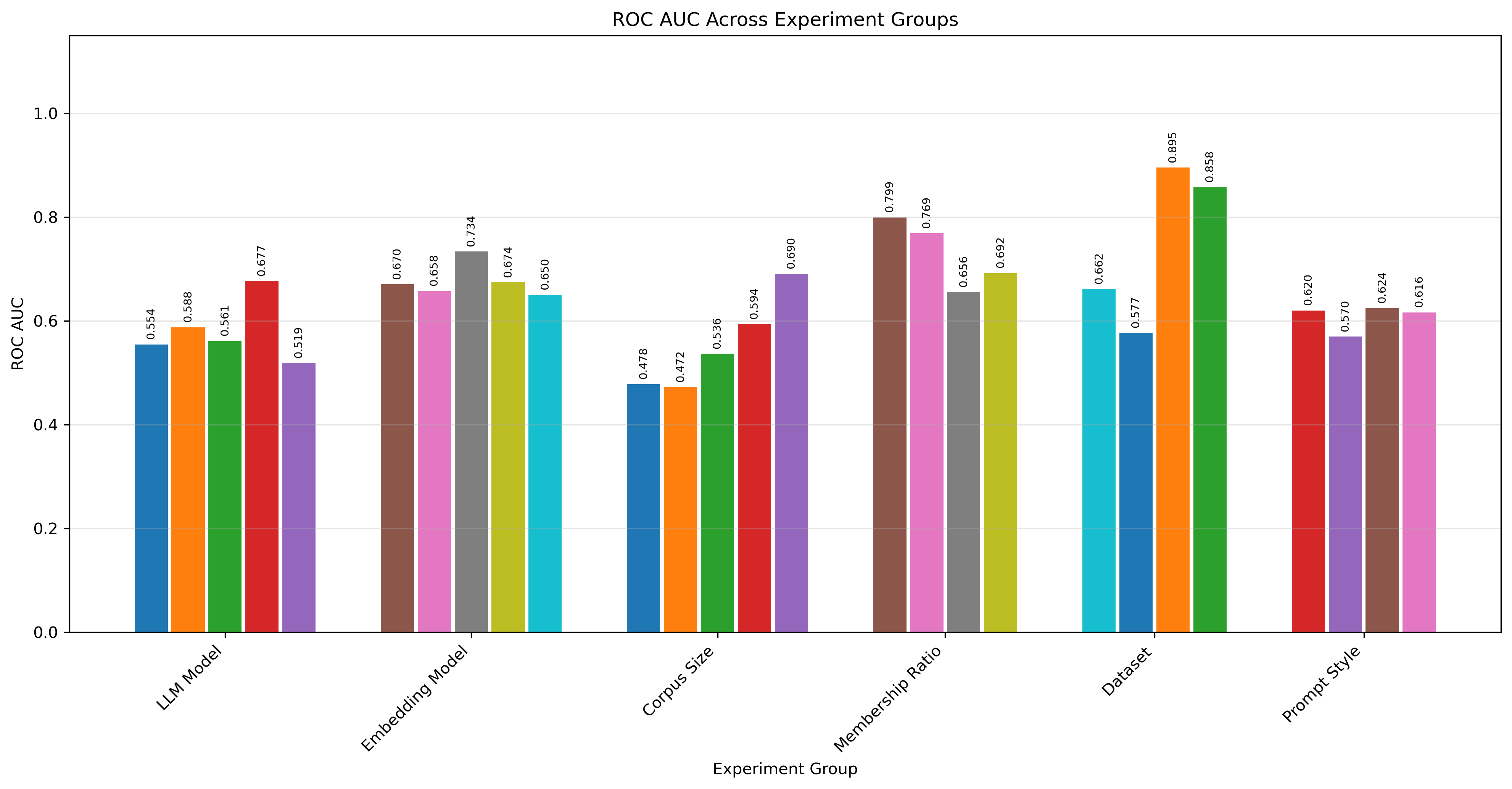}
\caption{Impact of the six evaluated factors on Membership Inference Attack (MIA) effectiveness, measured by ROC AUC.}
\label{fig:summary_barchart}
\end{figure*}

Figure~\ref{fig:summary_barchart} summarizes the impact of the six factors on membership inference effectiveness. Overall, dataset characteristics emerge as the most influential factor, followed by corpus size and membership ratio, whereas LLM choice, embedding model, and prompt style exhibit comparatively limited effects. These findings suggest that MIA vulnerability in RAG systems is driven primarily by the composition and characteristics of the retrieval data rather than by the underlying model or prompting configuration. However, the absence of confidence intervals and the generally modest discrimination achieved under several configurations limit the generalizability of these observations.

\subsubsection{AIA Ablation Summary}

\begin{figure*}[htbp]
    \centering
    \includegraphics[width=6in]{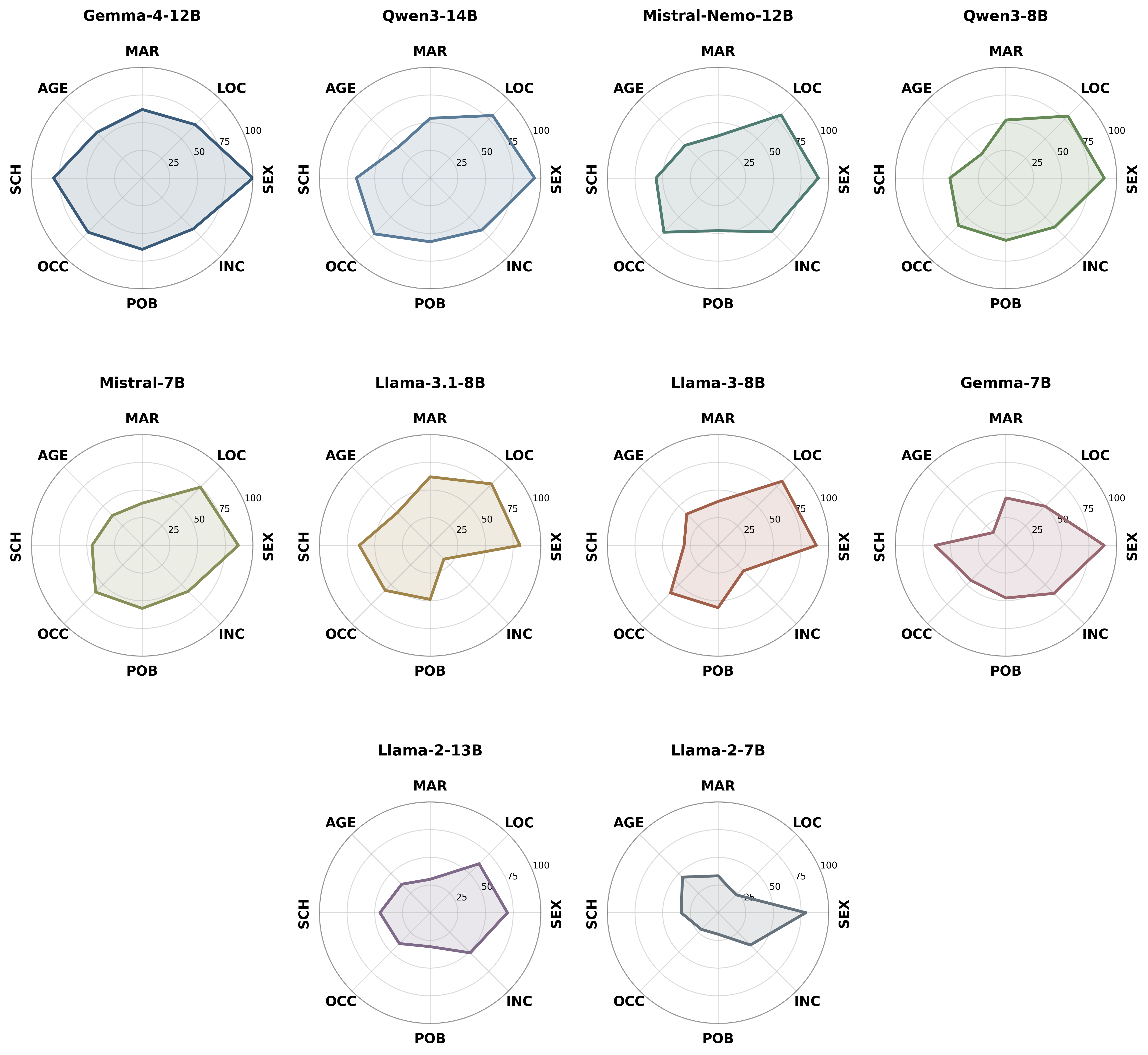}

    \vspace{0.3em}

    \begin{tabular}{llllllll}
        \hline
        \textbf{MAR} & \textbf{LOC} & \textbf{SEX} & \textbf{INC} &
        \textbf{POB} & \textbf{OCC} & \textbf{SCH} & \textbf{AGE} \\
        Relationship Status & Location & Gender & Income Level &
        Birthplace & Occupation & Education & Age \\
        \hline
    \end{tabular}

    \caption{AIA Top-1 Accuracy in across models in inferring each personal attribute.}
    \label{fig:radar-accuracy}
\end{figure*}

The result of the attribute-inference experiments are shown per their Top-1 and Top-3 results in Figure~\ref{fig:overall-accuracy} and detailed per attribute in Figure~\ref{fig:radar-accuracy}. Across low-cardinality attributes like gender, all models are highly-performing with a top-1 mean of 86.8\% and no score below 75\%. 
AIA attempts become less successful on average when the cardinality of the attribute increases as in Age for example where it is shown in Figure~\ref{fig:radar-accuracy} to have the lowest success on average with a mean of 38.9\% across all models. This can be explained by the large range of guesses that the model has to choose from and the ambiguity of this feature in the text corpora.       

\subsubsection*{DEA Ablation Summary}

Based on the ablation study of DEA (Tables~\ref{tab:dea-verified} and~\ref{tab:dea-dup}), one can summarize the findings as follows:
\begin{enumerate}
\item \textbf{Query access is enough.} A blank email header recovers hundreds
  of real email addresses per run from the larger models; for emails, knowing
  real training text adds little.
\item \textbf{Scale drives leakage, architecture does not.} Larger models leak
  more, and a larger share of their output is real, under every prompt style;
  the two same-size pairs show no consistent architecture effect.
\item \textbf{Repeated values leak.} A value seen 50 or more times is
  recovered about 160 times more often than a value seen once.
\item \textbf{PII-formatted output is not leakage.} In the reference setting,
  27 of 1,632 email-like strings are real training data.
\end{enumerate}

\subsection{Ablation Analysis Per Factor}
\label{subsec:analysis}

In the following, we summarize the key factors that influence the privacy attacks considered in this work and identify common and attack-specific trends in LLM privacy risks. 

\subsubsection{Model Scale Impact.} 
Model scale emerges as a key factor influencing LLM privacy vulnerability across the studied attacks. For DEA, leakage grows with model size, and faster than size. In the Pythia family, where size is the only difference between models, the number of real email addresses recovered per 5,000 prompts rises from 1 at 70M parameters to 263 at 1.4B; from 160M to 1.4B, a model about nine times larger leaks about 40 times more (Table~\ref{tab:dea-verified}). GPT-J-6B recovers the most under every prompt style. A larger share of their email-like output is also real, growing from 0.3\% at Pythia-70M to 7.5\% for GPT-J-6B. Larger models memorize more, and ordinary prompting is enough to reach it.

\begin{figure*}[htbp]
    \centering
    \includegraphics[width=6in]{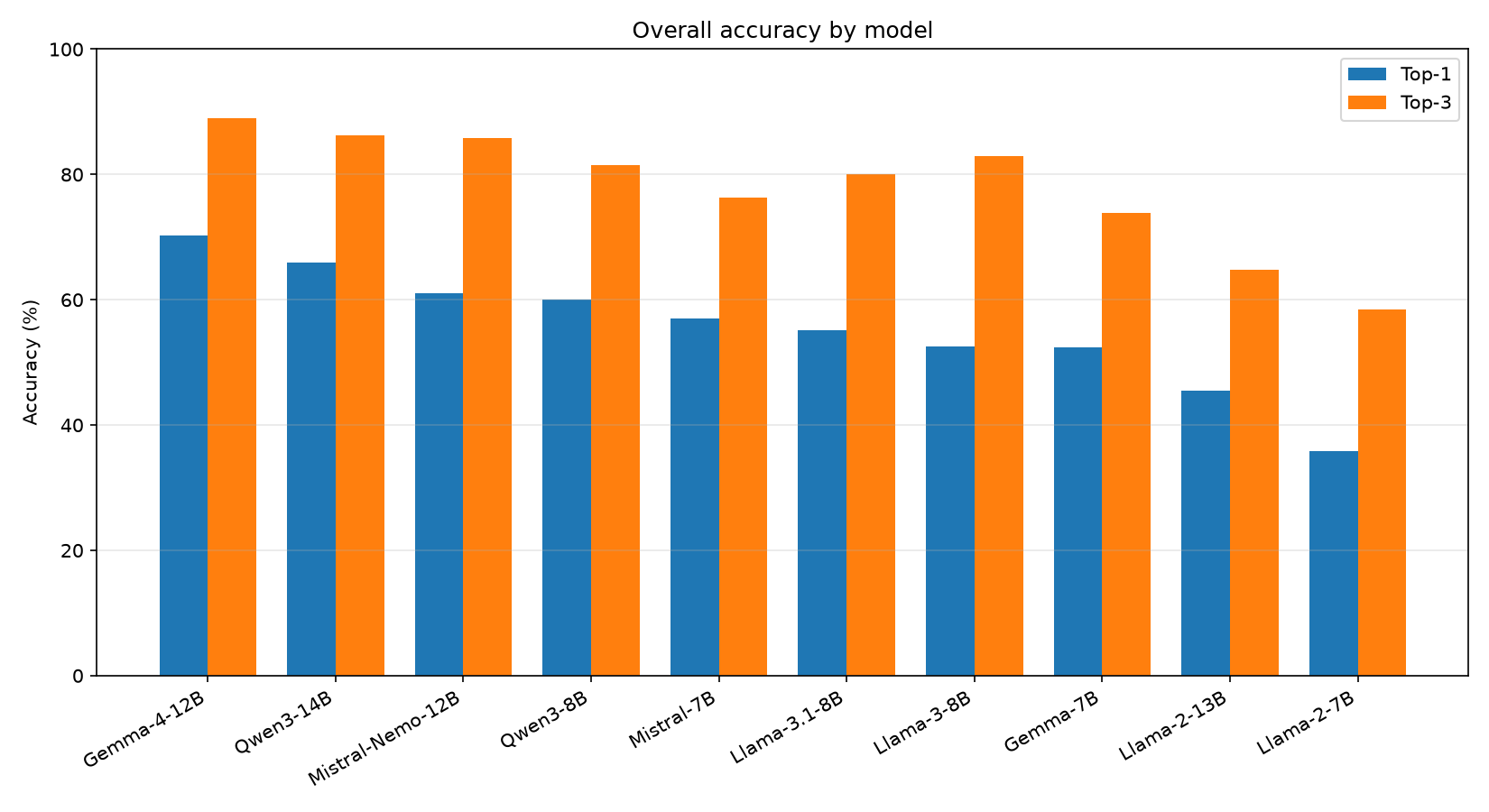}
    \caption{Overall Top-1 and Top-3 Accuracy across the models used in the AIA experiments.}
    \label{fig:overall-accuracy}
\end{figure*}

A similar positive scaling effect is observed for AIA. As shown in Figure~\ref{fig:parameter-accuracy} and Table~\ref{fig:overall-accuracy}, increasing model size from LLaMA-2-7B to LLaMA-2-13B and from Qwen-3-8B to Qwen-3--14B leads to a substantial improvement in inference accuracy.
This indicates that larger models are more effective at capturing and exploiting latent attribute correlations, thereby strengthening inference capabilities. From an adversarial perspective, model scale therefore constitutes a critical factor in maximizing attribute inference success.

For MIAs, evaluated through MBMIA (Appendix~\ref{sec:mbmia}), the impact of model scale remains inconclusive. As shown in Table~\ref{tab:mbmia_model_scale} and Figure~\ref{fig:rocauc-llmscale-mnmia}, Llama models generally achieve higher attack effectiveness than GPT models, particularly in terms of F1-score. However, this difference cannot be attributed solely to model size, as the models also differ in architecture and training characteristics. More importantly, comparisons within the same model family reveal no consistent improvement with scale: Llama-3.1-8B outperforms Llama-3.1-70B in F1-score, while GPT-4o-mini outperforms GPT-4o. Furthermore, ROC AUC remains near its maximum across all configurations, indicating strong membership discrimination regardless of scale. Overall, these results suggest that model family may influence attack effectiveness, but increasing model size does not systematically improve MIA performance.

For BAs, model scale reveals a potential trade-off between attack effectiveness and stealthiness. Comparing TinyLlama (1.1B) and LLaMA2 (7B), the larger model achieves a higher attack success rate (89.47\% vs. 85.7\%) but lower clean accuracy (84.2\% vs. 98.7\%). These results suggest that increased model capacity may improve backdoor effectiveness at the expense of clean performance. However, further within-family comparisons are needed to establish whether this trade-off consistently results from model scaling.

Overall, this indicates that larger models are typically more effective at capturing and exploiting latent attribute correlations, thereby strengthening inference capabilities. From an adversarial perspective, model scale therefore constitutes a critical factor in maximizing attribute inference success.

\subsubsection{Model Architecture Impact.} 
In contrast to the model scale, the impact of model architecture on privacy attacks is more nuanced and highly dependent on the underlying attack mechanism. 

For the MIA, both S2MIA and MBMIA variants exhibit relatively limited sensitivity to architectural differences. As shown in ~Table~\ref{tab:llm} and Table~\ref{tab:mbmia_model_family}, the performance metrics remain within a narrow range across models, indicating that the underlying membership signal is largely preserved regardless of architectural design. 

For DEA, we compare two pairs of models of similar size from different families, Pythia and GPT-Neo (Table~\ref{tab:dea-verified}). The small pair gives nearly identical counts under every prompt style. In the large pair, Pythia is higher in most cells, but the two models also differ slightly in size and in training settings, so the gap cannot be assigned to architecture. DEA therefore shows no consistent architecture effect. Only models with a public training set can be checked this way, which limits how many architectures can be compared.

For the BA, architectural differences lead to clear variations in both attack effectiveness and stealthiness. As shown in Figures~\ref{fig:barPlot} and~\ref{fig:heatBa}, GPT-2 achieves the highest Attack Success Rate (ASR) (95.2\%), followed by LLaMA2\_7B (89.47\%), Mistral-7B (89.1\%), and TinyLlama\_1.1B (85.7\%), indicating that all architectures remain vulnerable but differ in their ability to internalize and activate backdoor triggers. 

For AIA, model architecture and training characteristics exhibit varying degrees of association with attack effectiveness. As shown in Table~\ref{tab:architecture-comp} and Figures~\ref{fig:tokens-accuracy}, \ref{fig:contextwindow-accuracy}, and \ref{fig:layers-accuracy}, pre-training data volume shows the strongest positive association with attribute inference accuracy, suggesting that exposure to larger training corpora may improve the model's ability to infer sensitive attributes from textual cues. Context window size also exhibits a positive association, although its practical impact remains uncertain, while model depth shows a weaker and less consistent relationship. Hence, model architecture appears to influence AIA effectiveness, but architectural characteristics alone are insufficient to explain the observed differences, suggesting that privacy vulnerability emerges from the interplay between model design and pre-training characteristics.

Overall, these results indicate that the impact of model architecture on privacy risks is attack-dependent. While MIA exhibits limited sensitivity to architectural differences and DEA shows no consistent architectural effect, BA reveals variations in attack success and clean performance across model families. For AIA, architectural and training characteristics are associated with inference accuracy, although their individual contributions remain difficult to isolate. These findings suggest that model architecture does not uniformly amplify privacy risks but influences different attack mechanisms in distinct ways.

\subsubsection{Training Data Scale Impact.} The effect of training data scale on privacy attacks exhibits a consistent but nuanced pattern. For S2MIA, increasing the corpus size leads to a clear improvement in attack performance. In particular, ROC-AUC increases monotonically from near-random performance at smaller corpus fractions (10–50\%) to significantly higher values at full scale (up to $0.690$), indicating that larger datasets strengthen the membership signal. This trend is further supported by improvements in accuracy and precision, confirming that inference-based attacks benefit from richer data distributions.

For DEA, the property of the training data that matters is repetition: how many times a value occurs in it. An email address seen once was recovered about one time in 600 (3 of 1,806); one seen 50 or more times was recovered about one time in four (1,191 of 4,422). The 20\% of addresses seen 14 or more times account for 83\% of all addresses recovered (Table~\ref{tab:dea-dup}). Repetition is what turns a training value into an extractable one.

\subsubsection{Prompt Characteristics Impact.} 

Prompt style influences privacy attack effectiveness differently depending on the attack mechanism. For S2MIA, its impact on membership discrimination is relatively limited, with ROC AUC ranging from 0.570 to 0.624 (Table~\ref{tab:prompt_style}). However, structured prompts such as RTOC and CRISPE improve PR AUC, whereas Chain-of-Thought (CoT) weakens discrimination, potentially by introducing additional text that dilutes the similarity signals exploited by the attack.

For DEA, prompt style has a more pronounced impact on the amount and type of information disclosed (Fig.~\ref{fig:dea-scale}). While generic prompts can extract substantial amounts of real training-data PII without prior knowledge, prompts containing real training-text prefixes recover more phone numbers for the largest model (180 against 76 per run for GPT-J-6B). Nevertheless, the increase in leakage with model scale remains consistent across prompting strategies.

Overall, prompt style affects privacy attacks through different mechanisms: it influences response similarity and membership discrimination in MIA, while shaping the amount and type of information extracted in DEA. Its impact is therefore attack-dependent, and more elaborate prompts do not necessarily yield more effective attacks. Prompt style was not independently evaluated for AIA and BA in this study.

\subsection{Magnitude of Privacy Attacks on LLMs}

\begin{table*}[t]
\centering
\caption{Summary of the magnitude of privacy attacks. DEA values are counts or shares of real values recovered, not accuracies (Section~\ref{sec:DEA}).}
\label{tab:allexps}
\scriptsize
\begin{tabular}{llll}
\toprule
\textbf{Attack} & \textbf{Factor} & \textbf{Metric} & \textbf{Range of accuracy}  \\
\midrule
\multirow{6}{*}{S2MIA}  & Dataset domain & ROC AUC & $0.662 \rightarrow 0.858$  \\
                         & Corpus Size & ROC AUC & $0.478 \rightarrow 0.690$  \\
                         & Membership Ratio & ROC AUC & $0.656 \rightarrow 0.799$  \\
                         & LLM Architecture & ROC AUC & $0.554 \rightarrow 0.677$  \\
                         & Embedding Model & ROC AUC & $0.670 \rightarrow 0.734$  \\
                         & Prompt Style & ROC AUC & $0.570 \rightarrow 0.624$  \\
                         \midrule
\multirow{7}{*}{MBMIA}   & Model Scale & ROC AUC & $0.99 \rightarrow 1.0$  \\
                         & Model Architecture & ROC AUC & $0.71 \rightarrow 0.99$  \\
                         & Retrieval Stack & ROC AUC & $0.99 \rightarrow 0.99$  \\
                         & Domain Retrieval & ROC AUC & $0.99 \rightarrow 0.99$  \\
                         & Mask Count & ROC AUC & $0.90 \rightarrow 1.0$  \\
                         & Threshold $\gamma$ & ROC AUC & $0.99 \rightarrow 0.99$  \\
                         & Retrieval Depth & ROC AUC & $0.99 \rightarrow 0.99$  \\
                         \midrule
\multirow{2}{*}{AIA}     & Model Scale & Accuracy & $0.37 \rightarrow 0.73$  \\
                         & Model Architecture & Accuracy & $0.5 \rightarrow 0.73$  \\
                         \midrule
\multirow{5}{*}{DEA}     & Model Scale (Pythia 70M $\rightarrow$ 1.4B) & Verified emails per run (count) & $1 \rightarrow 263$ \\
                         & Prompt Style (GPT-J-6B, 3 styles) & Verified emails per run (count) & $114 \rightarrow 413$ \\
                         & Model Architecture (2 same-size pairs) & Verified emails per run (count) & no consistent effect \\
                         & Training Data Repetition & Share of values recovered & $0.2\% \rightarrow 26.9\%$ \\
                         & Ground-truth check (reference setting) & Verified / extracted emails & $27 / 1{,}632$ \\
                         \midrule
\multirow{2}{*}{BA}      & Model architecture & Attack Success Rate (ASR) & $0.85 \rightarrow 0.95$  \\
                         & Model scale & Attack Success Rate (ASR) & $0.84 \rightarrow 0.95$  \\          
\bottomrule
\end{tabular}
\end{table*}

Table~\ref{tab:allexps} shows, for each privacy attacks reproduced in this paper, and for each factor, the range of accuracy of the attack. The results reported in the table must be interpreted in light of the objective of each privacy attack, as this objective fundamentally determines the achievable performance. 

Membership Inference Attacks (MIA) are formulated as a binary classification problem, where the adversary aims to decide whether a specific record is present in the RAG dataset or not. Because the task reduces to distinguishing between two classes (member vs. non-member), relatively high performance—especially in terms of ROC AUC—is expected when the underlying signal is sufficiently strong. In contrast, Attribute Inference Attacks (AIA) aim to predict latent user attributes (e.g., age, gender, location) from user-generated text. The achievable accuracy in this setting is inherently constrained by the cardinality and ambiguity of the attribute space: low-cardinality attributes such as gender are easier to infer, while higher-cardinality attributes such as age ranges, city, or occupation introduce significantly more uncertainty, naturally lowering overall accuracy. Data Extraction Attacks (DEA) are measured differently: we count the real email addresses and phone numbers recovered from the training data, not an accuracy, so the DEA values in the table are counts or shares and are not comparable with the ROC AUC, accuracy or ASR of the other attacks. Finally, Backdoor Attacks (BA) follow a different paradigm altogether: their objective is not to infer or reconstruct, but to trigger a specific malicious behavior embedded during training. Their effectiveness is therefore measured by the attack success rate, i.e., the proportion of outputs that match the attacker-defined target, which can be high when the trigger mechanism is reliably learned. Overall, the diversity in accuracy ranges across the table directly reflects the differing goals and inherent difficulty of each attack class, rather than differences in experimental setup alone. 

\subsubsection{MIA}
For Membership Inference Attacks (MIA), the two evaluated approaches exhibit substantially different effectiveness. S2MIA, which splits a target record into a query and a held-out portion and relies on response similarity, achieves moderate and variable discrimination. In contrast, MBMIA employs a more targeted strategy by selectively masking multiple parts of a document and measuring the model's ability to reconstruct them. This provides a stronger membership signal, particularly when the target document is retrieved as context, resulting in consistently near-perfect ROC AUC across experimental configurations. 

\subsubsection{AIA} 
For Attribute Inference Attacks (AIA), the relatively low accuracy values observed are primarily explained by the intrinsic difficulty of the objective. In AIA, the model must infer latent user attributes from indirect linguistic cues, which are often ambiguous, overlapping, and highly dependent on context, especially for attributes with large value spaces. This makes the task inherently uncertain and limits achievable accuracy. As a result, AIA operates under weaker and noisier signals than membership inference, which explains its lower and more variable performance. 

\subsubsection{DEA}

For DEA, attack magnitude cannot be expressed as classification accuracy but is measured by the amount of verified training-data PII recovered. Although most generated PII-like strings do not match the training data (only 27 of 1,632 email-like strings in the reference setting), larger models disclose hundreds of real email addresses per run. Across all experiments, 2,934 distinct PII values were recovered. Thus, even a relatively low extraction rate can represent substantial privacy leakage, as each recovered value corresponds to potentially sensitive personal information.

\subsubsection{BA}
For BA, attack effectiveness is consistently high, with Attack Success Rates (ASR) ranging from 85.7\% to 95.2\% across the evaluated models. Unlike MIA and AIA, which rely on uncertain inference signals, BA exploits malicious behaviors deliberately introduced during training, making successful activation more predictable. However, high ASR does not necessarily imply a stealthy attack, as clean accuracy varies substantially across models (78.9–98.7\%). These results indicate that backdoor attacks can be highly reliable while remaining difficult to detect when normal model performance is preserved.

Table~\ref{tab:allexps} also reveals that the factors driving attack effectiveness differ across privacy attacks. For S2MIA, dataset domain has the strongest influence on membership discrimination, reflecting the sensitivity of its question–answer approach to data characteristics. For MBMIA, model architecture exhibits the largest variation, although attack effectiveness remains generally high. For AIA, model scale shows the widest accuracy range, but its impact is not consistently positive. For DEA, model scale and training-data repetition are the main drivers of verified leakage. Finally, for BA, model architecture produces greater variation in attack success than model scale. Overall, these findings demonstrate that no single factor consistently dominates privacy vulnerability across attack types, reinforcing the need for attack-specific evaluations.

\subsection{Practical Recommendations}
\label{subsec:recommendations}
Based on the analysis in Section~\ref{subsec:analysis}, we derive practical guidelines to help practitioners assess, for a given deployment scenario, which privacy attack an LLM is most vulnerable to.

\begin{enumerate}
\item Prioritize defenses against mask-based membership inference (MBMIA).
MBMIA achieves near-ceiling performance and is highly sensitive to model architecture, making it the most reliable membership signal. Systems using powerful models and RAG are therefore particularly exposed to membership leakage. Practitioners should treat this as a primary risk and apply defenses such as limiting verbatim retrieval, adding noise, and auditing reconstruction behavior.

\item Be cautious with semantic-based attacks (S2MIA) in domain-specific datasets.
S2MIA performance varies significantly with dataset domain and structure, indicating that some datasets leak more than others. Homogeneous or highly structured corpora increase vulnerability. Practitioners should ensure data diversity and avoid relying solely on prompt-level defenses.

\item AIA is harder but potentially damaging.
Although its accuracy is lower due to the difficulty of the task, AIA can directly expose sensitive personal information, and it improves with model scale. Practitioners should enforce strict access control, monitoring, and rate limiting, as stronger attackers can significantly increase its impact.

\item Remove repeated personal data before training, and re-check it when scaling up (DEA).
Extraction concentrates on values that recur in the training data: the 20\% of Enron email addresses seen 14 or more times account for 83\% of the addresses we recovered, and an address seen 50 or more times is recovered about 160 times more often than one seen once. Before training, practitioners should count how often each email address and phone number occurs in the corpus and remove or mask those that recur, in particular contact details copied into signatures, footers and quoted replies. This audit should be repeated whenever a larger model is trained on the same data, because leakage grows faster than model size: in the Pythia family, real addresses recovered per run rise from 6 at 160M to 263 at 1.4B parameters, so data that is safe for a small model may not be safe for a large one. Query-side filtering and rate limiting alone help little, since a blank email header, with no knowledge of the training data, recovers hundreds of real addresses per run from the larger models. Finally, extraction audits should check outputs against the training data and state which copy of the data and which count (distinct values or occurrences) they use: in our reference setting, only 27 of 1,632 email-like strings were real.

\item Treat backdoor attacks as a critical and systematic threat.
Backdoor attacks achieve high and stable success rates by triggering predefined behaviors. They are especially dangerous in fine-tuning and RAG settings, where poisoned inputs may be introduced. 
\end{enumerate}

\section{Conclusion}
\label{sec-conc}

In this paper, we presented a unified analysis of four major privacy attacks against LLMs (MIA, AIA, DEA, and BA) through a structured ablation study. Our findings reveal that privacy vulnerabilities are attack-dependent: MIA is particularly sensitive to dataset characteristics and attack strategy, AIA is influenced by model and training characteristics, DEA increases with model scale and training-data repetition, and BA exhibits consistently high success rates.

These findings provide practical guidance for privacy-aware LLM deployment. In particular, practitioners should prioritize membership leakage assessments in RAG systems, minimize repeated personal information in training data, monitor attribute inference risks, and secure fine-tuning pipelines against backdoor attacks. More broadly, privacy safeguards should be tailored to the dominant threats of each deployment scenario rather than applied uniformly.

Future work will investigate interactions between these factors, evaluate the effectiveness of targeted defenses, and extend the analysis to larger models and real-world deployment settings.
\\

\section*{AI Use Statement}
We used generative AI tools primarily to improve the wording, grammar, and readability of the manuscript, to assist with literature screening and checking relevant prior work, and to support the polishing of plots and figures. The scientific contributions of the paper, including the proposed unified threat model, common notation and framework, selection and design of the ablation factors, experimental methodology, and interpretation of the findings, were developed entirely by the authors. All AI-assisted edits, literature suggestions, and visualization changes were reviewed and verified by the authors. The authors take full responsibility for the accuracy of the text, experiments, figures, references, and conclusions.

\section*{Ethics Statement}
This study evaluates privacy and security risks in large language models using publicly available research datasets and locally executed models. No human participants were recruited or contacted, and no new personal data were collected. The MIA experiments use established benchmark datasets and evaluate record membership in controlled RAG settings; they do not involve newly collected personal data or interaction with human participants. The AIA experiments use fully synthetic conversations. For the DEA experiments, we use the publicly available Enron corpus and its occurrence in The Pile solely to verify whether model outputs correspond to real training-data identifiers. PII comparisons are performed using hashed values, generated model outputs containing PII are not released, and real-text prefixes containing personal information are excluded from the released artifacts.  
The BA experiments are conducted in a controlled research setting using established benchmarks, and harmful outputs are not reproduced in the paper. All experiments were performed for privacy and security evaluation, with the goal of understanding and mitigating risks rather than enabling misuse. The released code and artifacts are limited to what is necessary for reproducibility and do not include sensitive personal information.

\section*{Reproducibility Statement}
We release the code, experimental configurations, dataset-processing scripts, and non-sensitive result files needed to reproduce the experiments reported in this paper. The repository documents the evaluated models, datasets, decoding parameters, retrieval settings, software environment, hardware configuration, and random seeds where applicable. Per-run or per-seed results are provided when available to support independent verification of the reported analyses and figures.  
To preserve privacy, artifacts containing real PII, generated outputs containing personal information, and real-text prefixes containing personal information are not released. These omissions do not affect the release of the experimental code, configurations, aggregate results, or scripts used to reproduce the reported analyses. All released materials are available at the repository referenced in the paper.

%Bibliography
\bibliographystyle{unsrt}  
\bibliography{references}

\newpage 

\appendix
\section{An additional MIA Attack}
\subsection{Mask-Based Membership Inference Attack (MBMIA)}
\label{sec:mbmia}
Similarly to the previous attack (S2MIA), MBMIA~\cite{liu2025maskbased} has the same goal. That is, given a target document $d$, an adversary aims to determine whether $d$ was included in the retriever ($R$)'s indexed corpus (``Member'') or not (``Non-Member'') of the RAG Dataset $\mathcal{D}$. 
Given a user query $Q$ and a RAG dataset $\mathcal{D}$, the retriever $R$ returns the top-$K$ documents
\begin{equation}
    D_{K} = R(Q,\mathcal{D},K),
\end{equation}
and the generator then produces a response using the system prompt $s$, a document-specific probe $Q_d$, and the retrieved context:
\begin{equation}
    r = \mathcal{M}(s \oplus Q_d \oplus D_{K}).
\end{equation}

\textbf{Threat Model.} We assume the black-box setting used by in~\cite{liu2025maskbased}: the attacker cannot access the index contents, retriever $R$ internals, embedding vectors, or model $\mathcal{M}$ weights, but can repeatedly query the deployed RAG system and observe only its textual outputs $D_{K}$. The goal is to build a binary membership inference classifier (BMIC) that outputs the likelihood of $d \in \mathcal{D}$:

\begin{equation}
    P(d \in \mathcal{D}) = \mathrm{BMIC}(d, r).
\end{equation}

\subsubsection{Mechanism}
\label{subsubsec:mbmia-mechanism}
% The original paper is \textit{Mask-based Membership Inference Attacks for Retrieval-Augmented Generation} by Liu, Zhang, and Long \cite{liu2025maskbased}. 
Our implementation keeps the core mechanism as Liu et al.\cite{liu2025maskbased}, but our experiments differ in two important ways. First, we explicitly compare a hosted GPT model and local open-weight models. Second, we extend beyond the original three datasets and probe additional system factors such as retrieval depth and index size.

The main goal of MBMIA is to assess the efficiency of the MIA using masks. The basic idea is to mask terms in the query based on a target document $d$ which can efficiently indicate whether $d \in \mathcal{D}$. More precisely, MBMIA works because it masks terms that are intentionally difficult to reconstruct from general background knowledge alone. If $d$ itself is retrieved, the generator receives the exact lexical evidence needed to recover those terms. If the target document is absent ($d \notin \mathcal{D}$), the model may still produce plausible paraphrases, broad concepts, or approximate answers, but it should fail more often on exact reconstruction \cite{liu2025maskbased}. That exact-match asymmetry is what makes mask accuracy useful as a membership signal.

The end-to-end attack pipeline is as follows.
\begin{enumerate}
    \item \textbf{Proxy scoring.} A proxy language model is used to estimate which words or phrases are hardest to predict from local context alone; the masking policy targets those difficult units rather than random words \cite{liu2025maskbased}. This is better described as a \textit{difficulty ranking} step than as a generic perplexity computation.
    \item \textbf{Mask generation.} The selected terms are replaced with placeholders such as \texttt{[Mask\_1]}, while preserving an answer key for exact-match evaluation. Let $m$ denotes the number of generated masks.
    \item \textbf{Probe submission.} The masked document is sent to the RAG system as the document-specific probe $Q_d$.
    \item \textbf{Mask reconstruction.} The generator is instructed to recover each masked item in a strict output format.
    \item \textbf{Membership inference.} Mask reconstruction accuracy is treated as the membership score, and a threshold based on $\gamma$ converts that score into a member/non-member decision.
\end{enumerate}

\begin{figure}[t]
\centering
\begin{tikzpicture}[
    font=\small,
    >=Latex,
    node distance=5mm,
    box/.style={
        draw,
        rounded corners=2pt,
        align=center,
        minimum height=7mm,
        text width=6.6cm,
        fill=blue!6,
        inner sep=3pt
    },
    retr/.style={
        draw,
        rounded corners=2pt,
        align=center,
        minimum height=7mm,
        text width=6.6cm,
        fill=green!10,
        inner sep=3pt
    },
    note/.style={
        draw,
        rounded corners=2pt,
        align=left,
        text width=6.6cm,
        fill=orange!10,
        inner sep=4pt
    },
    flow/.style={->, thick},
    helper/.style={->, thick, dashed}
]

% main vertical pipeline
\node[box] (doc) {\textbf{Target document}\\$d$};

\node[box, below=of doc] (proxy)
{\textbf{1) Proxy scoring}\\rank hard terms};

\node[box, below=of proxy]
(mask) {\textbf{2) Mask generation}\\create $M$ masks};

\node[box, below=of mask]
(probe) {\textbf{3) Submit masked query}\\document as $Q_d$};

\node[retr, below=of probe]
(retriever) {\textbf{4) Retrieve context}\\top-$K$ documents};

\node[box, below=of retriever]
(recon) {\textbf{5) LLM reconstruction}\\predict $[\mathrm{Mask}_i]$};

\node[box, below=of recon]
(infer) {\textbf{6) Membership decision}\\mask accuracy + $\gamma$};

% heuristics below
\node[note, below=7mm of infer]
(heur) {\textbf{Heuristics from Liu et al.}\\
fragmented-word handling \quad $\bullet$\quad
misspelling correction \quad $\bullet$\quad
adjacency filtering};

% main arrows
\draw[flow] (doc) -- (proxy);
\draw[flow] (proxy) -- (mask);
\draw[flow] (mask) -- (probe);
\draw[flow] (probe) -- (retriever);
\draw[flow] (retriever) -- (recon);
\draw[flow] (recon) -- (infer);

% helper arrows
\draw[helper] (heur.north east) to[out=110,in=-70] (mask.south east);
\draw[helper] (heur.north west) to[out=70,in=-110] (recon.south west);

\end{tikzpicture}

\caption{Revised self-contained view of the MBA pipeline adapted from Liu et al.~\cite{liu2025maskbased}. The workflow proceeds from document scoring and mask construction to masked-query submission, retrieval of top-$K$ context, reconstruction of masked terms, and final membership decision using mask accuracy and threshold $\gamma$. The stabilizing heuristics support both mask construction and reconstruction reliability.}

\label{fig:mba_framework_revised}
\end{figure}

\subsubsection{Factors Affecting the Attack}

To rigorously evaluate the MIA attack performance, we vary the LLM model ($\mathcal{M}$), Retrieval method $R$, embedding model, RAG dataset ($\mathcal{D}$) domain, and the attack-specific hyperparameters $m$ (number of masks), $\gamma$ (threshold fraction of correct masks needed), and $K$ (number of retrieved documents from $\mathcal{D}$).

\subsubsection{Metrics}

We use the following metrics:

\begin{itemize}
    \item \textbf{Mask reconstruction accuracy (Mask Accuracy).} For each target document, we first generate a masked probe and then ask the RAG system to fill in the missing terms. Mask Accuracy is the fraction of masked positions whose reconstructed value exactly matches the ground-truth answer key for a given target document. A higher Mask Accuracy means that the generator recovered more of the masked lexical items required by the probe.

    \item \textbf{Attack success (ROC AUC).} We treat Mask Accuracy as a continuous membership score and compute ROC AUC over member and non-member examples. AUC measures ranking quality: 1.0 means perfect separation, whereas 0.5 is random guessing.

    \item \textbf{F1-score.} F1 captures the quality of the final binary member/non-member decision after thresholding. It is especially useful for analyzing the choice of $\gamma$, because $\gamma$ changes the operating point even when the underlying ranking (AUC) remains almost unchanged.
   
    \item \textbf{Retrieval Recall.} Retrieval Recall measures whether the target document appears among the top-$K$ retrieved contexts when the document is truly a member. A \textit{retrieval failure} occurs when the true member document is absent from the top-K retrieved context. A \textit{generation failure} occurs when the true member document is retrieved successfully, but the response still does not reconstruct enough masked items to meet the membership criterion; in our implementation, this means retrieval\_hit = 1 and Mask Accuracy < $\gamma$.
\end{itemize}

\subsection{Experimental Results and Analysis}

We now present the experimental evaluation of the MBMIA under a controlled ablation framework. Each experiment isolates a specific factor while keeping all other components fixed to baseline settings.

The baseline configuration consists of a balanced dataset with 50\% member and 50\% non-member samples, using GPT-4o-mini as the generator, FAISS as the retriever, BGE-small as the embedding model, $m = 10$, $K = 5$, and $\gamma = 0.5$. Across all experiments, Retrieval Recall remains consistently equal to 1.0, indicating that the retriever successfully returns the target document when it is present. Therefore, differences in attack performance primarily originate from generation behavior rather than retrieval failures.

\subsubsection*{Experiment~\ref{sec:mbmia}.A --- Model Scale Impact}

We begin by comparing the models' scale across two LLMs, namely GPT and Llama variants. Table~\ref{tab:mbmia_model_scale} and Figures~\ref{fig:rocauc-llmscale-mnmia} and~\ref{fig:mbmia_model_scale_f1} summarize the results.

\begin{table}[!t]
\centering
\scriptsize
\caption{MBMIA performance across different model scales.}
\label{tab:mbmia_model_scale}
\begin{tabular}{lccc}
\toprule
Model & ROC AUC & F1 & Recall \\
\midrule
Llama 3.1-8B & 1.0000 & 0.9434 & 1.0000 \\
Llama 3.1-70B & 0.9994 & 0.8929 & 1.0000 \\
GPT-4o & 0.9980 & 0.7874 & 1.0000 \\
GPT-4o-mini & 0.9950 & 0.8772 & 1.0000 \\
\bottomrule
\end{tabular}
\end{table}

\begin{figure}[!t]
\centering
\scriptsize
\begin{tikzpicture}
\begin{axis}[
    ybar,
    bar width=18pt,
    width=\linewidth,
    height=7cm,
    ymin=0.75, ymax=1.02,
    ylabel={ROC AUC},
    symbolic x coords={Llama3.1-8B,Llama3.1-70B,GPT-4o,GPT-4o-mini},
    xtick=data,
    xticklabel style={rotate=20, anchor=east},
    nodes near coords,
    every node near coord/.append style={font=\scriptsize, rotate=90, anchor=west},
    enlarge x limits=0.12,
    grid=major
]
\addplot coordinates {
    (Llama3.1-8B,1.0000)
    (Llama3.1-70B,0.9994)
    (GPT-4o,0.9980)
    (GPT-4o-mini,0.9950)
};
\end{axis}
\end{tikzpicture}
\caption{ROC AUC across model scales.}
\label{fig:rocauc-llmscale-mnmia}
\end{figure}

\begin{figure}[!t]
\centering
\scriptsize
\begin{tikzpicture}
\begin{axis}[
    ybar,
    bar width=18pt,
    width=\linewidth,
    height=7cm,
    ymin=0.70, ymax=1.00,
    ylabel={F1-score},
    symbolic x coords={Llama3.1-8B,Llama3.1-70B,GPT-4o,GPT-4o-mini},
    xtick=data,
    xticklabel style={rotate=20, anchor=east},
    nodes near coords,
    every node near coord/.append style={font=\scriptsize, rotate=90, anchor=west},
    enlarge x limits=0.12,
    grid=major
]
\addplot coordinates {
    (Llama3.1-8B,0.9434)
    (Llama3.1-70B,0.8929)
    (GPT-4o,0.7874)
    (GPT-4o-mini,0.8772)
};
\end{axis}
\end{tikzpicture}
\caption{F1-score across different model scales.}
\label{fig:mbmia_model_scale_f1}
\end{figure}

Our implementation already included GPT-4o-mini as the hosted-model reference point. The present revision expands that comparison by adding a dedicated model-scale study rather than introducing GPT for the first time. The results confirm that attack performance remains near-ceiling across all strong models, with differences primarily reflected in F1 due to output formatting and reconstruction precision.

\subsubsection*{Experiment~\ref{sec:mbmia}.B --- Model Architecture Impact}

We then evaluate the impact of four LLMs on the attack performance while keeping all other factors fixed. Table~\ref{tab:mbmia_model_family} shows the results.

\begin{table}[!t]
\centering
\scriptsize
\caption{MBMIA performance across model families on HealthcareMagic.}
\label{tab:mbmia_model_family}
\begin{tabular}{lccc}
\toprule
Model & ROC AUC & F1 & Recall \\
\midrule
Llama 3 & 0.9980 & 0.9524 & 1.0000 \\
GPT-4o-mini & 0.9934 & 0.8850 & 1.0000 \\
Phi-3 & 0.9104 & 0.7179 & 1.0000 \\
Mistral & 0.7144 & 0.7711 & 1.0000 \\
\bottomrule
\end{tabular}
\end{table}

Because Retrieval Recall is 1.0 for all model-family runs, the observed AUC differences are unlikely to originate from retrieval. A more defensible interpretation is that the attack is most sensitive to generation-side behavior under strict exact-match scoring: some models follow the mask-reconstruction format more consistently, return shorter and less noisy answers, and more often reproduce the exact lexical items required by the answer key. In that setting, stronger attack performance should be read as cleaner exposure of the membership signal under the probe protocol, not as a blanket statement that one model family is universally more vulnerable in every RAG deployment.

\begin{table}
\centering
\scriptsize
\caption{LLM ($\mathcal{M}$) comparison on HealthcareMagic ($m=10$, $K=5$, FAISS, BGE-small, $\gamma=0.5$).}
\label{tab:model_family}
\begin{tabular}{lccc}
\toprule
Model & AUC & F1 & Retrieval Recall \\
\midrule
Llama~3 & 0.9980 & 0.9524 & 1.0000 \\
GPT-4o-mini & 0.9962 & 0.8696 & 1.0000 \\
Phi-3 & 0.9266 & 0.7342 & 1.0000 \\
Mistral & 0.7144 & 0.7711 & 1.0000 \\
\bottomrule
\end{tabular}
\end{table}

\subsubsection*{Experiment~\ref{sec:mbmia}.C --- Retrieval Stack Impact}

To isolate the effect of the retrieval pipeline, we fix the generator (GPT-4o-mini) and vary retriever and embedding combinations. Table~\ref{tab:mbmia_retrieval_stack} and Figure~\ref{fig:mbmia_domain_control_auc} show minimal variation across configurations.

\begin{table}[!t]
\centering
\scriptsize
\caption{MBMIA performance across retrieval stacks on HealthcareMagic.}
\label{tab:mbmia_retrieval_stack}
\begin{tabular}{lcc}
\toprule
Retriever + Embedding & ROC AUC & F1 \\
\midrule
FAISS + all-MiniLM-L6-v2 & 0.9958 & 0.8621 \\
BM25 + all-MiniLM-L6-v2 & 0.9950 & 0.8696 \\
BM25 + BGE-small-en-v1.5 & 0.9942 & 0.8621 \\
FAISS + BGE-small-en-v1.5 & 0.9932 & 0.8850 \\
\bottomrule
\end{tabular}
\end{table}

This indicates that, under perfect-recall conditions, the retrieval stack contributes marginally to attack performance compared to the generator. The attack is therefore not retrieval-limited in this setup.

\subsubsection*{Experiment~\ref{sec:mbmia}.D --- Domain vs Retrieval Stack Impact}

We evaluate whether dataset domain or retrieval stack has a stronger influence on attack performance. Table~\ref{tab:mbmia_domain_control} and Figure~\ref{fig:mbmia_domain_control_auc} summarize the within-dataset AUC ranges.

\begin{table}[!t]
\centering
\scriptsize
\caption{Within-dataset AUC variation across retrieval stacks.}
\label{tab:mbmia_domain_control}
\begin{tabular}{lccc}
\toprule
Dataset & Mean AUC & Min AUC & Max AUC \\
\midrule
ArXiv & 0.9889 & 0.9886 & 0.9890 \\
FiQA & 0.9978 & 0.9964 & 0.9992 \\
HealthcareMagic & 0.9946 & 0.9940 & 0.9950 \\
\bottomrule
\end{tabular}
\end{table}

\begin{figure}[!t]
\centering
\scriptsize
\begin{tikzpicture}
\begin{axis}[
    ybar,
    bar width=20pt,
    width=0.85\linewidth,
    height=7cm,
    ymin=0.985, ymax=1.001,
    ylabel={Mean ROC AUC},
    symbolic x coords={ArXiv,FiQA,HealthcareMagic},
    xtick=data,
    nodes near coords,
    every node near coord/.append style={font=\scriptsize},
    enlarge x limits=0.2,
    grid=major
]
\addplot coordinates {
    (ArXiv,0.9889)
    (FiQA,0.9978)
    (HealthcareMagic,0.9946)
};
\end{axis}
\end{tikzpicture}
\caption{Mean ROC AUC across datasets under controlled retrieval-stack comparison.}
\label{fig:mbmia_domain_control_auc}
\end{figure}

The current cross-dataset results show that strong MBA performance is not limited to medical text when the retrieval stack is fixed, but they do not by themselves isolate domain effects from retriever or embedding effects. The appropriate conclusion is therefore modest: multiple domains appear vulnerable under the baseline stack, but the relative contribution of domain versus retrieval-stack choice must be established through controlled ablation.

\subsubsection*{Experiment~\ref{sec:mbmia}.E --- Mask Count ($m$) Impact}

We vary the number of masks $m$ while keeping all other factors fixed. Table~\ref{tab:mbmia_mask_count} and Figure~\ref{fig:mbmia_mask_count_auc} show a strong improvement as $m$ increases from 3 to 15, followed by saturation.

\begin{table}[!t]
\centering
\scriptsize
\caption{Effect of the number of masks $m$ on MBMIA performance.}
\label{tab:mbmia_mask_count}
\begin{tabular}{ccc}
\toprule
$m$ & ROC AUC & F1 \\
\midrule
3 & 0.9032 & 0.8214 \\
5 & 0.9426 & 0.8522 \\
10 & 0.9932 & 0.8696 \\
15 & 1.0000 & 0.9434 \\
20 & 1.0000 & 0.9174 \\
\bottomrule
\end{tabular}
\end{table}

\begin{figure}[!t]
\centering
\scriptsize
\begin{tikzpicture}
\begin{axis}[
    width=0.9\linewidth,
    height=7cm,
    ymin=0.88, ymax=1.02,
    xlabel={Number of masks $m$},
    ylabel={ROC AUC},
    xtick={3,5,10,15,20},
    grid=major,
    nodes near coords,
    every node near coord/.append style={font=\scriptsize, anchor=south}
]
\addplot[
    mark=*,
] coordinates {
    (3,0.9032)
    (5,0.9426)
    (10,0.9932)
    (15,1.0000)
    (20,1.0000)
};
\end{axis}
\end{tikzpicture}
\caption{ROC AUC as a function of the number of masks $m$.}
\label{fig:mbmia_mask_count_auc}
\end{figure}

Small values of $m$ lead to unstable scores due to high sensitivity to individual reconstruction errors. Increasing $m$ stabilizes the signal by averaging over more masked positions, leading to near-perfect separation at higher values. However, this saturation should be interpreted cautiously, as it likely reflects the controlled evaluation setup rather than a universally optimal configuration.
% \ckar{Didn't get this interpretation.}

\subsubsection*{Experiment~\ref{sec:mbmia}.F --- Threshold ($\gamma$) Impact}

We analyze the effect of the decision threshold $\gamma$ on attack performance. Table~\ref{tab:mbmia_gamma} and Figure~\ref{fig:mbmia_gamma_f1} show that while ROC AUC remains nearly constant across threshold choices, F1 varies substantially.

\begin{table}[!t]
\centering
\scriptsize
\caption{Effect of decision threshold $\gamma$ on MBMIA performance.}
\label{tab:mbmia_gamma}
\begin{tabular}{ccc}
\toprule
$\gamma$ & ROC AUC & F1 \\
\midrule
0.3 & 0.9950 & 0.7634 \\
0.5 & 0.9932 & 0.8772 \\
0.7 & 0.9904 & 0.9709 \\
0.9 & 0.9956 & 0.8636 \\
\bottomrule
\end{tabular}
\end{table}

\begin{figure}[!t]
\centering
\scriptsize
\begin{tikzpicture}
\begin{axis}[
    width=0.9\linewidth,
    height=7cm,
    ymin=0.70, ymax=1.00,
    xlabel={Threshold $\gamma$},
    ylabel={F1-score},
    xtick={0.3,0.5,0.7,0.9},
    grid=major,
    nodes near coords,
    every node near coord/.append style={font=\scriptsize, anchor=south}
]
\addplot[
    mark=*,
] coordinates {
    (0.3,0.7634)
    (0.5,0.8772)
    (0.7,0.9709)
    (0.9,0.8636)
};
\end{axis}
\end{tikzpicture}
\caption{F1-score as a function of threshold $\gamma$.}
\label{fig:mbmia_gamma_f1}
\end{figure}

This confirms that $\gamma$ primarily controls the operating point of the classifier rather than its ranking ability. Therefore, $\gamma$ should be interpreted as a decision calibration parameter rather than a factor affecting inherent attack separability.
% \ckar{Didn't get this interpretation.}

\subsubsection*{Experiment~\ref{sec:mbmia} G: Retrieval Depth ($K$) Impact}

We vary the number of retrieved documents $K$ while keeping all other factors fixed. Table~\ref{tab:mbmia_retrieval_depth} 
% and Figure~\ref{fig:mbmia_retrieval_depth_auc}
shows that ROC AUC remains consistently high for all tested values.

\begin{table}[!t]
\centering
\scriptsize
\caption{Effect of retrieval depth $K$ on MBMIA performance.}
\label{tab:mbmia_retrieval_depth}
\begin{tabular}{ccc}
\toprule
$K$ & ROC AUC & F1 \\
\midrule
1 & 0.9992 & 0.8850 \\
3 & 0.9956 & 0.8850 \\
5 & 0.9970 & 0.8772 \\
10 & 0.9938 & 0.8696 \\
\bottomrule
\end{tabular}
\end{table}

\begin{table*}
\centering
\caption{Representative best and worst configurations from the revised experiment set.}
\label{tab:best_runs}
\scriptsize
\begin{tabular}{llllccc}
\toprule
\textbf{Model} & \textbf{Dataset} & \textbf{Retriever} & \textbf{Embedding} & \textbf{$M$/$K$} & \textbf{AUC} & \textbf{F1} \\
\midrule
GPT-4o-mini & HealthcareMagic & FAISS & BGE-small & $15/5$ & 1.0000 & 0.9434 \\
GPT-4o-mini & HealthcareMagic & FAISS & BGE-small & $20/5$ & 1.0000 & 0.9174 \\
Llama~3 & HealthcareMagic & FAISS & BGE-small & $10/5$ & 0.9980 & 0.9524 \\
GPT-4o-mini & FiQA & FAISS & all-MiniLM-L6-v2 & $10/5$ & 0.9992 & 0.9009 \\
Mistral & HealthcareMagic & FAISS & BGE-small & $10/5$ & 0.7144 & 0.7711 \\
\bottomrule
\end{tabular}
\end{table*}

This suggests that once the correct document is retrieved, additional context provides limited benefit and may introduce minor noise in reconstruction.

\subsubsection*{Ablation Study: Best vs. Worst Configurations}

The best observed configuration in the current study is $\mathcal{M}=$GPT-4o-mini, $\mathcal{D}=$HealthcareMagic with $m=15$, $R=$FAISS, embedding=BGE-small, $K=5$, and $\gamma=0.5$, reaching $AUC = 1.0000$ and $F1 = 0.9434$. The worst model-family configuration under matched conditions is Mistral on HealthcareMagic with $AUC = 0.7144$. These extremes again reinforce that, once retrieval recall is fixed at $1.0$, the main source of variation is the generator's behavior under the exact-match reconstruction prompt.

\end{document}